\documentclass[aps,pra,a4paper,reprint,twocolumn,floatfix,superscriptaddress,notitlepage,usenames,dvipsnames,svgnames,accepted=2022-06-16,table,nofootinbib,longbibliography]{quantumarticle}
\pdfoutput=1
\usepackage[utf8]{inputenc}
\UseRawInputEncoding
\usepackage[T1]{fontenc}
\usepackage{graphicx}
\usepackage{grffile}
\usepackage{wrapfig}
\usepackage{rotating}
\usepackage[normalem]{ulem}
\usepackage{amsmath}
\usepackage{textcomp}
\usepackage{amssymb}
\usepackage{capt-of}
\usepackage{verbatim}

\usepackage{parskip}
\usepackage{mathrsfs}
\usepackage[margin= 0.75in]{geometry}
\usepackage[braket, qm]{qcircuit}

\usepackage[english]{babel}
\usepackage{letltxmacro}
\LetLtxMacro{\ORIGselectlanguage}{\selectlanguage}
\makeatletter
\DeclareRobustCommand{\selectlanguage}[1]{%
  \@ifundefined{alias@\string#1}
    {\ORIGselectlanguage{#1}}
    {\begingroup\edef\x{\endgroup
      \noexpand\ORIGselectlanguage{\@nameuse{alias@#1}}}\x}%
}
\newcommand{\definelanguagealias}[2]{%
  \@namedef{alias@#1}{#2}%
}
\makeatother
\definelanguagealias{en}{english}

\usepackage{bbm}
\usepackage{etoolbox}
\usepackage{tcolorbox} 
\usepackage{fleqn} 
\usepackage{tikz}
\usepackage{amsthm}
\usepackage{amssymb}
\usetikzlibrary{arrows.meta}
\usepackage{mathtools}

\DeclarePairedDelimiter\ceil{\lceil}{\rceil}
\DeclarePairedDelimiter\floor{\lfloor}{\rfloor}

\newcommand{\prlsection}[1]{{\em {#1}.---~}}

\usepackage{bm}
\usepackage{dsfont}
\usepackage[font=small,labelfont=bf,justification=justified,format=plain]{caption}
\usepackage{subcaption}

\usepackage{thmtools}
\usepackage{thm-restate}
\newtheorem{thm}{Theorem}
\newtheorem{prop}[thm]{Proposition}

\definecolor{blueviolet}{rgb}{0.54, 0.17, 0.89}
\definecolor{mycitecolor}{rgb}{0,0.7,0.1}
\usepackage{hyperref}
\hypersetup{
 pdfauthor={Namit Anand},
 pdftitle={Information Scrambling at Finite Temperature},
 pdflang={English},colorlinks=true,linkcolor=blueviolet,citecolor=teal,urlcolor=Periwinkle} 
\usepackage[capitalise]{cleveref}

\begin{document}
\title{BROTOCs and Quantum Information Scrambling at Finite Temperature}

\author{Namit Anand}
\email [e-mail: ]{namitana@usc.edu}

\author{Paolo Zanardi}
\email [e-mail: ]{zanardi@usc.edu}

\affiliation{Department of Physics and Astronomy, and Center for Quantum Information Science and Technology, University of Southern California, Los Angeles, California 90089-0484, USA}

\date{June 16, 2022}

\begin{abstract}
Out-of-time-ordered correlators (OTOCs) have been extensively studied
in recent years as a diagnostic of quantum information scrambling. In this
paper, we study quantum information-theoretic aspects of the 
\textit{regularized} finite-temperature OTOC. We introduce analytical
results for the \textit{bipartite regularized} OTOC (BROTOC): the regularized OTOC averaged over random
unitaries supported over a bipartition. We show that the BROTOC has several interesting properties, for example, it quantifies the purity of the associated thermofield double state and the ``operator purity'' of the analytically continued time-evolution
operator. At infinite-temperature, it reduces to one minus the
operator entanglement of the time-evolution operator. In the
zero-temperature limit and for nondegenerate Hamiltonians, the BROTOC probes the
groundstate entanglement. By computing long-time averages, we show that the equilibration value of the
BROTOC is intimately related to eigenstate entanglement. Finally, we numerically study the equilibration value of the BROTOC for various physically relevant Hamiltonian models and comment on its ability to distinguish integrable and chaotic dynamics.
\end{abstract}
\maketitle

\section{Introduction}
\label{sec:intro}
The thermalization of closed quantum systems has been a long standing
puzzle in theoretical physics
\cite{srednicki_chaos_1994,
  rigol_thermalization_2008, dalessio_quantum_2016,
  borgonovi_quantum_2016}. Recently, the notion of ``information
scrambling'' as the underlying mechanism for thermalization has gained
prominence. The idea is that complex quantum systems quickly
disseminate localized information through the (nonlocal) degrees of
freedom, making it inaccessible to any \textit{local} probes to the system. The
information is not lost, since the global evolution is still unitary,
rather, it is encoded in nonlocal correlations across the system. A quantification of this dynamical phenomena has initated a rich discussion surrounding operator growth \cite{PhysRevX.8.021013,PhysRevX.8.021014,PhysRevX.8.031058,PhysRevX.8.031057,PhysRevB.98.220303,PhysRevX.8.041019,parker_universal_2019}, eigenstate thermalization hypothesis (ETH) \cite{murthy_bounds_2019}, quantum chaos \cite{maldacena_bound_2016, xu_does_2020}, among others; see also Refs. \cite{swingle_unscrambling_2018,xu_swingle_tutorial_2022} for a recent review. One of the central objects in this quantification are the so-called out-of-time-ordered correlators (OTOCs). The OTOC is usually defined as a four point function with unusual time-ordering \cite{larkin_quasiclassical_1969, kitaev_simple_2015},
\begin{align}
\label{eq:otoc-unregularized-defn}
F_{\beta}(t) := \operatorname{Tr}\left[ W^{\dagger}_{t} V^{\dagger} W_{t} V \rho_{\beta} \right],
\end{align}
where \(W_{t}:= U^{\dagger}_{t} W U_{t}\) is the Heisenberg-evolved operator and \(\rho_{\beta} = \exp \left[ -\beta H \right] / \mathcal{Z}(\beta)\) is the Gibbs state at inverse temperature \(\beta\) with \(\mathcal{Z}(\beta) := \operatorname{Tr}\left[ \exp \left[ -\beta H \right]  \right]\). An intimately related quantity to the above OTOC is the following norm of the commutator,
\begin{align}
C_{\beta}(t) &:= \frac{1}{2} \operatorname{Tr}\left[ \left[ W_{t},V \right]^{\dagger} [W_{t},V] \rho_{\beta} \right] \nonumber\\
&= \frac{1}{2} \left\Vert \left[ W_{t},V \right] \sqrt{\rho_{\beta}} \right\Vert_{2}^{2}.
\end{align}
Here we have used the Hilbert-Schmidt norm \(\left\Vert \cdot \right\Vert_{2}\), which originates from the (Hilbert-Schmidt) inner product \(\left\langle A,B \right\rangle:= \operatorname{Tr}\left[ A^{\dagger}B \right]\). The two quantities are related via the simple formula,
\begin{align}
C_{\beta}(t) = 1 - \mathrm{Re} F_{\beta}(t).
\end{align}
Therefore, the growth of the norm of the commutator is associated to the decay of the OTOCs.

The idea behind using the norm of the commutator to quantify scrambling is the following: let \(V\) and \(W\)
be two local operators that initially commute (for example, consider
local operators on two different sites of a quantum spin-chain). Under
Heisenberg time-evolution, the support of \(W_{t}\) grows and after
a transient period, it will start noncommuting with the operator \(V\)
and one can utilize the commutator \(C_{\beta}(t)\) to quantify this
growth. Intuitively, if the Hamiltonian of this system is local, then
Leib-Robinson type bounds can provide an estimate for the time it
takes for the growth of this commutator
\cite{liebFiniteGroupVelocity, hastingsSpectralGapExponential2006, bravyi_lieb-robinson_2006}.

Understanding quantitatively, the scrambling of quantum information has lead
to a plethora of theoretical insights \cite{maldacena_bound_2016,
  cotler_black_2017, hosur_chaos_2016, PhysRevX.8.021013,PhysRevX.8.021014,PhysRevX.8.031058,PhysRevX.8.031057,PhysRevB.98.220303,PhysRevX.8.041019,parker_universal_2019,
  murthy_bounds_2019}. This was swiftly followed by several
state-of-the-art experimental investigations
\cite{mi2021information,braumuller2021probing,Wei_2018,
  li_measuring_2017,nie2019detecting,Nie_2020,garttner_measuring_2017,Joshi_2020,Meier_2019,Chen_2020}. Furthermore,
several works have now elucidated quantum information theoretic
aspects underlying the OTOC, for example, by connecting it to Loschmidt Echo \cite{yan_information_2020}, operator entanglement and entropy production \cite{styliaris_information_2021, zanardi_information_2021-1}, quantum coherence \cite{anand_quantum_2020}, entropic uncertainty relations \cite{halpern_entropic_2019}, among others.

In Refs. \cite{yan_information_2020, styliaris_information_2021}, the
authors defined a ``bipartite OTOC,'' obtained by averaging the
infinite-temperature OTOC uniformly over local random unitaries
supported on a bipartition. In Ref. \cite{styliaris_information_2021}, this bipartite OTOC was shown to have the
following operational interpretations: (i) it is exactly the
\textit{operator entanglement}
\cite{zanardi_entanglement_2001-1,PhysRevA.66.044303} of the dynamical
unitary \(U_{t}\), (ii) it connects in a simple way to the entangling
power \cite{zanardiEntanglingPowerQuantum2000} of the dynamical
unitary \(U_{t}\), (iii) it is exactly equal to the average linear entropy
production power of the reduced dynamics, among others. Furthermore,
several of these connections were generalized to the case of
open-system dynamics in Ref. \cite{zanardi_information_2021-1}, where,
in particular, a competition between information scrambling and
environmental decoherence was uncovered \cite{PRXQuantum.2.010306}.

Unfortunately, as we move away from the infinite-temperature assumption, the
connections unveiled in Ref. \cite{styliaris_information_2021} do not
carry over their operational aspects anymore. For example, a
straightforward generation to the finite temperature case, say, by
using the OTOC as defined in \cref{eq:otoc-unregularized-defn} fails
to retain the operator entanglement or entropy production
connection. Not all is lost, however, as it is the \textit{regularized} OTOC \cite{maldacena_bound_2016} that naturally lends
itself to these operational connections. Elucidating this connection
is the key technical contribution of this work. For ease of
readability, the proofs of key Propositions appear in the Appendix.

\section{Preliminaries}
\label{sec:preliminaries}
The OTOC introduced in \cref{eq:otoc-unregularized-defn} will hereafter be referred to as the \textit{unregularized} OTOC. In contrast, the \textit{regularized} (or symmetric) OTOC is defined as \cite{maldacena_bound_2016},
\begin{align}
\label{eq:regularized-otoc-defn}
F_{\beta}^{(r)}(t):= \operatorname{Tr}\left[ W^{\dagger}_{t} y V^{\dagger} y W_{t} y V y \right] \text{ with } y^{4} = \rho_{\beta}.
\end{align}
Equivalently,
\begin{align}
\label{eq:regularized-otoc-defn-x}
F_{\beta}^{(r)}(t)= \frac{1}{\mathcal{Z}(\beta)} \operatorname{Tr}\left[ W^{\dagger}_{t} x V^{\dagger} x W_{t} x V x \right],
\end{align}
with \(x = \exp \left[ -\beta H/4 \right]\). We also define the
associated disconnected correlator \cite{maldacena_bound_2016},
\begin{align}
F_{\beta}^{(d)}(t):= \operatorname{Tr}\left[ \sqrt{\rho_{\beta}} W^{\dagger}_{t} \sqrt{\rho_{\beta}} W_{t} \right]  \operatorname{Tr}\left[ \sqrt{\rho_{\beta}} V^{\dagger} \sqrt{\rho_{\beta}} V \right].
\end{align}

In Ref. \cite{maldacena_bound_2016}, a bound on the growth of the
correlator \(F^{(d)}_{\beta}(t) - F^{(r)}_{\beta}(t)\) was obtained
under certain assumptions as
\begin{align}
\frac{\partial  }{\partial t} \log \left( F_{\beta}^{(d)}(t) - F_{\beta}^{(r)}(t) \right) \leq \frac{2 \pi}{\beta}.
\end{align}
We also refer the reader to Ref. \cite{murthy_bounds_2019} the same bound was derived for systems satisfying ETH, along with some extra assumptions. In this work we focus on the quantity \(F^{(d)}_{\beta}(t) - F^{(r)}_{\beta}(t)\) arising from this bound and connect it to operational, quantum information-theoretic quantities. Notice that, for a time-independent Hamiltonian, the disconnected correlator \(F^{(d)}_{\beta}(t)\) is time independent (by using the commutation of \(\left[ y^{2}, U_{t} \right]\) and the cyclicity of trace). Therefore, we can define, \(F^{d}_{\beta} \equiv F^{d}_{\beta}(t) = \operatorname{Tr}\left[ y^{2} W^{\dagger} y^{2} W \right]  \operatorname{Tr}\left[ y^{2} V^{\dagger} y^{2} V \right]\).

Following Ref. \cite{styliaris_information_2021}, we will consider the
following setup: let \(\mathcal{H}_{AB}=\mathcal{H}_{A} \otimes
\mathcal{H}_{B} \cong \mathbb{C}^{d_{A}} \otimes \mathbb{C}^{d_{B}}\)
be a bipartition of the Hilbert space. Define as
\(\mathcal{U}(\mathcal{H}_{A(B)})\), the unitary group over
\(\mathcal{H}_{A(B)}\). We want to understand the qualitative and
quantitative features of the OTOC for a \textit{generic} choice of
local operators \(V\) and \(W\). Therefore, we average over unitary
operators supported on the bipartition \(A|B\). We define the
bipartite averaged, \emph{unregularized} OTOC (hereafter,
bipartite unregularized OTOC) as \cite{styliaris_information_2021}
\begin{align}
\label{eq:def-botoc}
G_{\beta}(t) := \mathbb{E}_{V_A,W_B} \left[C_{\beta}(t)\right],
\end{align}
where, \(V_{A} = V \otimes \mathbb{I}_{B}, W_{B} = \mathbb{I}_{A} \otimes W\), with
\(V\in\mathcal{U}(\mathcal{H}_A),\,W\in\mathcal{U}(\mathcal{H}_B),\)
and \(\mathbb{E}_{V,W}\left[\bullet\right]:= \int_{\mathrm{Haar}} dV\,
dW \left[\bullet\right]\) denotes Haar-averaging over the standard
uniform measure over \(\mathcal{U}(\mathcal{H}_{A(B)})\) \cite{watrousTheoryQuantumInformation2018}. In Ref. \cite{styliaris_information_2021} it
was shown that one can analytically perform the Haar averages to
obtain the following expression,
\begin{align}
G_{\beta}(t) = 1-\frac{1}{d} \operatorname{Re} \operatorname{Tr}\left(\left(\rho_{\beta} \otimes \mathbb{I}_{A^{\prime} B^{\prime}}\right) U_{t}^{\dagger \otimes 2} \mathbb{S}_{A A^{\prime}} U_{t}^{\otimes 2} \mathbb{S}_{A A^{\prime}}\right),
\end{align}
where \(\mathbb{S}_{AA'}\) is the operator that swaps the \(A
\leftrightarrow A'\) spaces in \(\mathcal{H}_{A} \otimes
\mathcal{H}_{B} \otimes \mathcal{H}_{A'} \otimes
\mathcal{H}_{B'}\). This equation represents the finite temperature version of the unregularized bipartite OTOC. For \(\beta=0\), this is the operator entanglement of the time evolution operator \(U_t\) as will be discussed shortly. However, for \(\beta \neq 0\), it does not have a clear quantum information-theoretic correspondence. 

Ref. \cite{styliaris_information_2021} studied \(G_{\beta}(t)\) in extensive detail at \(\beta=0\). Here, we will contrast the
dynamical behavior of the bipartite unregularized OTOC with that of the regularized case, which we are now ready to introduce. Performing bipartite averages in a similar way for the regularized case, we have,
\begin{align}
&  N_{\beta}(t) := G^{(d)}_{\beta} - G^{(r)}_{\beta}(t), \\
  & \text{with }  G^{(d)}_{\beta}:= \mathbb{E}_{V_{A},W_{B}} \left[ F^{(d)}_{\beta}\right],\\
  & \text{and } G^{(r)}_{\beta}(t):= \mathbb{E}_{V_{A},W_{B}} \left[ F^{(r)}_{\beta}(t) \right].
\end{align}
In the next section, we will discuss information-theoretic aspects of
these quantities. We also refer the reader to Refs. \cite{parker_universal_2019, tsuji_bound_2018,
  foini_eigenstate_2019, vijay_finite-temperature_2018,
  sahu_information_2020,PhysRevB.98.205124} for a discussion of various information scrambling/operator growth aspects of the regularized versus unregularized OTOCs.

\prlsection{Operator Schmidt decomposition} We take a small detour to
remind the reader a few key facts about operator entanglement before
delving into out main results. Given a pure quantum state in a bipartite Hilbert space, \(| \psi \rangle
\in \mathcal{H} \cong
\mathcal{H}_{A}\otimes \mathcal{H}_{B}\), there exists a Schmidt
decomposition of this state \cite{nielsen_quantum_2010},
\begin{align}
  | \psi \rangle = \sum\limits_{j=1}^{r} \sqrt{\lambda_{j}} | j_{A} \rangle \otimes | j_{B} \rangle.
\end{align}
Here, \(\{ \lambda_{j} \}_{j}\) are nonnegative coefficients with \(r =
\min (d_{A},d_{B})\) the Schmidt rank and \(\{ | j_{A} \rangle \}_{j=1}^{d_{A}},
\{| j_{B} \rangle \}_{j=1}^{d_{B}}\) bases for the subsystems \(A,B\),
respectively. The Schmidt coefficients can be used to compute various entanglement measures for the bipartite state \(| \psi
\rangle\) \cite{plenioIntroductionEntanglementMeasures2005}. The key idea behind Schmidt decomposition is to use the singular value decomposition for the matrix of coefficients obtained from expressing the state \(| \psi \rangle\) with respect to local orthonormal bases. In fact, one can generalize this idea to the operator space. Namely, consider bipartite operators, i.e., elements of \(\mathcal{L}(\mathcal{H}_{A} \otimes \mathcal{H}_{B})\), then we can define an \textit{operator Schmidt decomposition} \cite{zanardi_entanglement_2001-1,lupo2008bipartite,aniello2009relation}. Formally, given a bipartite operator \(X \in \mathcal{L}(\mathcal{H}_{A} \otimes \mathcal{H}_{B})\), there exist orthogonal bases \(\{ U_{j} \}_{j=1}^{d_{A}^{2}}\) and \(\{ W_{j} \}_{j=1}^{d_{B}^{2}}\) for \(\mathcal{L}(\mathcal{H}_{A}), \mathcal{L}(\mathcal{H}_{B})\), respectively, such that \(\left\langle U_{j},U_{k} \right\rangle = d_{A} \delta_{jk}\) and \(\left\langle W_{j}, W_{k} \right\rangle = d_{B} \delta_{jk}\). Moreover,
\begin{align}
X = \sum\limits_{j=1}^{\tilde{r}} \sqrt{\lambda_{j}} U_{j} \otimes W_{j}.
\end{align}
The coefficients \(\{ \lambda_{j} \}_{j}\) are nonnegative and are
called the operator Schmidt coefficients and \(\tilde{r} = \min \{
d_{A}^{2},d_{B}^{2} \}\) is the operator Schmidt rank. In fact, the operator entanglement of a unitary introduced in Ref. \cite{zanardi_entanglement_2001-1} is exactly the linear entropy of the probability vector \(\vec{p} = (\lambda_{1}, \lambda_{2}, \cdots, \lambda_{\tilde{r}})\) arising from the operator Schmidt coefficients. A key result obtained in Ref. \cite{zanardi_entanglement_2001-1} was that the operator entanglement of a unitary operator can be equivalently expressed as,
\begin{align}
E_{\mathrm{op}} \left( U \right) = 1 - \frac{1}{d^{2}} \operatorname{Tr}\left[ \mathbb{S}_{AA'} U^{\otimes 2} \mathbb{S}_{AA'} U^{\dagger \otimes 2}\right].
\end{align}
In a similar spirit, we define the \textit{operator purity} of a
linear operator as the purity of the probability vector \(\vec{p}\) obtained following the operator Schmidt decomposition. Namely,
\begin{align}
\mathcal{P}_{\mathrm{op}}(X) := \frac{1}{\left\Vert X \right\Vert_{2}^{4}} \operatorname{Tr}\left[ \mathbb{S}_{AA'} X^{\otimes 2} \mathbb{S}_{AA'} X^{\dagger \otimes 2}\right],
\end{align}
where we have explicitly introduced the normalization \(\left\Vert X \right\Vert_{2}^{4}\) for arbitrary operators (it is equal to \(d^{2}\) for unitaries which recovers the previous formula above).

Lastly, we remind the reader that, for unitary dynamics, information scrambling is usually quantified via the OTOCs, the operator entanglement of the time-evolution operator \(U_t\), and the quantum mutual information \cite{hosur_chaos_2016}. Our work, in particular, focuses extensively on the interplay between OTOCs and operator entanglement.
\section{Main results}
\label{sec:main}
\subsection{Operator entanglement}
\label{sec:operator-entanglement}
Our first result is to bring \(N_{\beta}(t)\) into an exact analytical
form. We introduce some notation first. Let \(\mathcal{P}_{\chi}(\rho):=
\left\Vert \rho_{\chi} \right\Vert_{2}^{2}\) be the squared \(2\)-norm of the operator \(\rho_{\chi}\) with \(\rho_{\chi}:= \operatorname{Tr}_{\overline{\chi}}\left[ \rho \right]\), \(\chi = \{ A,B \}\), and \(\overline{\chi}\) the complement of \(\chi\). If \(\rho\) is a quantum state then \(\mathcal{P}_{\chi}(\rho)\) is the purity across the \(A|B\) partition.\\

\begin{restatable}{prop}{regularizedotocmain}
\label{prop:regularized-otoc-main}
The regularized bipartite OTOC at finite temperature is
\begin{align}
  N_{\beta}(t) &= \frac{1}{d} \mathcal{P}_{A}(\sqrt{\rho_{\beta}}) \mathcal{P}_{B}(\sqrt{\rho_{\beta}}) \\ \nonumber
  &- \frac{1}{d \mathcal{Z}_{\beta}} \operatorname{Tr}\left[ \mathbb{S}_{AA'} \mathcal{U}_{\beta,t}^{\otimes 2} \left( \mathbb{S}_{AA'} \right) \right],
\end{align}
where, \(\mathcal{U}_{\beta,t}:= \mathcal{V}_{\beta} \circ
\mathcal{U}_{t}\) with \(\mathcal{V}_{\beta}(X):= \exp \left[ -\beta
  H/4 \right] X \exp \left[ -\beta H/4 \right] \) the imaginary time-evolution,
\(\mathcal{U}_{t}(X):= U^{\dagger}_{t} X U_{t}\) the real
time-evolution, and \(U_{t} = \exp
\left[ -iHt \right] \) the usual time-evolution operator.
\end{restatable}

Let us note a few simple things about this result:
(i) at infinite temperature (\(\beta = 0\)), this reduces to the
operator entanglement of the time evolution operator
\cite{styliaris_information_2021} \(G_{\beta=0}(t)\). The
equilibration value of this quantity was used to distinguish various
integrable and chaotic models, see
Refs. \cite{styliaris_information_2021, zhou_operator_2017} for more
details. (ii) In quantum information theory
\cite{nielsen_quantum_2010}, the most general description of the
dynamics of a quantum system is given by a completely positive (CP) and
trace-nonincreasing map, also called a \textit{quantum
  operation}. Furthermore, if the evolution is not only trace
non-increasing, rather, trace-preserving (TP), then such dynamical
maps are called  \textit{quantum channels}. In the Appendix, we show
that \(\mathcal{U}_{\beta,t}\) is a quantum operation. Moreover, the
second term, \(G^{(r)}_{\beta}(t)\) is real and proportional to the operator purity of
\(U_{\beta}(t):= \exp \left[ -(\beta - it) H/4 \right] \), the
analytically continued time-evolution operator, with \(\mathcal{Z}(\beta/2)^{2}/ \left( d
  \mathcal{Z}(\beta) \right)\) as the proportionality factor. (iii) The following simple upper bound holds
for the BROTOC: \(N_{\beta}(t) = G_{\beta}^{(d)} -
G_{\beta}^{(r)}(t) \leq G_{\beta}^{(d)} \leq \mathcal{Z}(\beta/2)^{4}/
\left( d \mathcal{Z}(\beta)^{2} \right)\). (iv) For a non-entangling
Hamiltonian, we have, \(N_{\beta}=0 ~~\forall \beta\). Namely, if \(H =
H_{A} + H_{B}\), then a simple calculation reveals that, \(G^{(d)}_{\beta} =
\frac{\mathcal{Z}(\beta/2)^{2}}{d \mathcal{Z}(\beta)} =
G^{(r)}_{\beta}\) and therefore, \(N_{\beta}(t)\) is identically
vanishing at all \(\beta\). Of course, the fact that at \(\beta = 0\),
\(N_{\beta}=0\) also follows from the connection to operator
entanglement \cite{styliaris_information_2021}.

We emphasize that, although several previous works have
focussed on understanding the growth of local OTOCs
in terms of  Lieb-Robinson bounds \cite{kukuljan_weak_2017-1,
  lin_out--time-ordered_2018,PhysRevB.100.064305,PhysRevB.98.144304,chen2016universal,PhysRevResearch.2.043234},
this analysis \textit{does not} apply to our bipartite OTOCs
(regularized or unregularized). The key distinction here is that, our
averaging is over observables supported on a bipartition \(A|B\) of
the \textit{entire} system (and not some subset of the total Hilbert
space). As a result, even if one of the subsystems is local its
complement is (highly) nonlocal. As a result, Lieb-Robinson type bounds
are not necessarily useful in understanding the growth of this quantity.

\subsection{BROTOC, thermofield double, and the
  spectral form factor}
In this section, we focus on the quantum operation
\(\mathcal{U}_{\beta,t}\), the operator purity of which is quantified
by the connected
BROTOC. We will show that the map \(\mathcal{U}_{\beta,t}\)
contains information about both spectral and eigenstate signatures of
quantum chaos \cite{haake_quantum_2010, guhr_random-matrix_1998, mehta_random_2004, dalessio_quantum_2016}. In particular, we will establish its relation to the
spectral form factor (SFF) \cite{berry1985semiclassical} and the
thermofield double state (TDS) \cite{takahashi_thermo_1996}. Recently,
several works have elucidated the ability of the TDS to probe scrambling and
quantum chaos
\cite{cotler_black_2017, Dyer2017thermofield,del_campo_scrambling_2017,PhysRevLett.115.211601}. In its
simplest form, the TDS corresponds to a ``canonical'' purification of the Gibbs state \(\rho_{\beta} = \exp \left[ -\beta H
\right] / \mathcal{Z}(\beta)\). Given the connections to scrambling
and chaos, the ability to experimentally prepare TDS allows
us to directly probe these properties; see for e.g.,
Refs. \cite{PhysRevLett.123.220502,PhysRevA.100.032107,zhu2020generation,Cottrell_2019,PhysRevResearch.2.013254}
for a discussion about how to prepare such states on a quantum
computer.

More
formally, let \(| \Gamma \rangle :=
\sum\limits_{j=1}^{d} | j \rangle | j \rangle\) be the
\textit{unnormalized} maximally entangled vector in
\(\mathcal{H}^{\otimes 2}\), then, the TDS is defined as,
\begin{align}
\label{eq:thermofield-double-purification}
  | \psi(\beta) \rangle := \left( \sqrt{\rho_{\beta}} \otimes \mathbb{I} \right) | \Gamma \rangle.
\end{align}
By construction, \(| \psi(\beta) \rangle \in
\mathcal{H}^{\otimes 2}\) and tracing out either subsystem gives
us back the original Gibbs state. For simplicity, consider a nondegenerate Hamiltonian with a spectral
decomposition \(H = \sum\limits_{j=1}^{d} E_{j} | j \rangle \langle j|\), then, by considering the \(| \Gamma \rangle\) matrix expressed
with respect to the Hamiltonian eigenbasis, we have,
\begin{align}
\label{eq:thermofield-double-energy}
| \psi(\beta) \rangle = \frac{1}{\sqrt{\mathcal{Z}(\beta)}}\sum\limits_{j=1}^{d} \exp \left[ -\beta E_{j}/2 \right] | j \rangle | j \rangle.
\end{align}
Written in this form, it is easy to see that partial tracing either subsystem of \(|
\psi(\beta) \rangle\) generates the Gibbs state \(\rho_{\beta}\). In
Ref. \cite{del_campo_scrambling_2017}, the survival probability (or
Loschmidt Echo) of the
time-evolving TDS was related to the analytically continued partition function \cite{maldacena_bound_2016,
  cotler_black_2017,Dyer2017thermofield}. Namely, let the
time-evolved TDS be defined as \cite{takahashi_thermo_1996, del_campo_scrambling_2017},
\begin{align}
|\psi(\beta,t) \rangle &:= \left( U_{t} \otimes \mathbb{I} \right) |
  \psi(\beta) \rangle \nonumber\\
  &= \frac{1}{\sqrt{\mathcal{Z}(\beta)}}
\sum\limits_{j=1}^{d} \exp \left[ -(\beta/2 + it) E_{j} \right] | j
\rangle | j \rangle,
\end{align}
then its survival probability is
\begin{align}
\left| \left\langle \psi(\beta,0) | \psi(\beta,t) \right\rangle \right|^{2} = \frac{\left| \mathcal{Z}(\beta+it) \right|^{2}}{\mathcal{Z}(\beta)^{2}}.
\end{align}
This is clearly related to the two-point, analytically continued
SFF, which is defined as \cite{cotler_black_2017, cotler_chaos_2017},
\begin{align}
  \mathcal{R}_{2}(\beta,t):= \left\langle \left| \mathcal{Z}(\beta+it) \right|^{2} \right\rangle_{\mathrm{RMT}},
\end{align}
where \(\left\langle \cdots \right\rangle_{\mathrm{RMT}}\) denotes an
ensemble average, usually over a random matrix ensemble of
Hamiltonians \cite{guhr_random-matrix_1998}. 

We will now show that an analogous, though, not identical, result holds for the quantum
operation \(\mathcal{U}_{\beta,t}\). Namely, we will consider the
fidelity between the Choi-Jamiolkowski (CJ) matrix \cite{wildeClassicalQuantumShannon2016} corresponding to
\(\mathcal{U}_{\beta,t}\) and \(\mathcal{U}_{\beta,0}\) and show that
it is related to the two-point SFF. Recall that the Choi-Jamiolkowski isomorphism is an isomorphism between linear maps \(\mathcal{E}:
\mathcal{L}(\mathcal{H}) \rightarrow \mathcal{L}(\mathcal{K})\) to
matrices \(\rho_{\mathcal{E}} \in \mathcal{L}(\mathcal{H}) \otimes
\mathcal{L}(\mathcal{K})\) \cite{wildeClassicalQuantumShannon2016}. Let \(| \phi^{+} \rangle:=
\frac{1}{\sqrt{d}} | j \rangle | j \rangle\) be the
\textit{normalized} maximally entangled state in \(\mathcal{H}^{\otimes 2}\), then,
\begin{align}
  \rho_{\mathcal{E}}:= \mathcal{E} \otimes \mathcal{I} \left( | \phi^{+} \rangle \langle  \phi^{+} |  \right).
\end{align}
A linear map \(\mathcal{E}\) is CP \(\iff \rho_{\mathcal{E}} \geq
0\). Now, a simple calculation shows that the CJ matrix corresponding
to the quantum operation \(\mathcal{U}_{\beta,t}\) is,
\begin{align}
\rho_{\mathcal{U}_{\beta,t}} = \frac{\mathcal{Z}(\beta/2)}{d} | \psi(\beta/2,t) \rangle \langle  \psi(\beta/2,t) | .
\end{align}

To quantify
how close two pure quantum states are, we can compute the fidelity
\cite{nielsen_quantum_2010} between them. Recall that the
fidelity between two pure quantum states \(| \psi \rangle, | \phi \rangle\) is given as,
\begin{align}
F(| \psi \rangle,| \phi \rangle) = \left| \left\langle \psi | \phi \right\rangle \right|^{2},
\end{align}
with \(F(| \psi \rangle,| \phi \rangle) = 1
\iff | \psi \rangle  = | \phi \rangle\).
Since the Choi matrix \(\rho_{\mathcal{U}_{\beta,t}}\) is proportional to a pure-state
projector, the fidelity between the matrices
\(\rho_{\mathcal{U}_{\beta,t}}\) and \(\rho_{\mathcal{U}_{\beta,0}}\) can be defined as,
\begin{align}
  &F(\rho_{\mathcal{U}_{\beta,t}}, \rho_{\mathcal{U}_{\beta,0}}) \nonumber\\
 &\equiv \left( \frac{\mathcal{Z}(\beta/2)}{d} \right)^{2}  F\left(| \psi(\beta/2,t) \rangle, | \psi(\beta/2,0) \rangle \right) \nonumber\\
  & = \left( \frac{\mathcal{Z}(\beta/2)}{d} \right)^{2} \left| \left\langle \psi(\beta/2,t) | \psi(\beta/2,0) \right\rangle \right|^{2} \nonumber\\
  &= \frac{\mathcal{R}^{H}_{2}(\beta/2,t)}{d^{2}},
\end{align}
where \(\mathcal{R}_{2}^{H}\) is the two-point SFF \textit{before}
ensemble averaging \cite{cotler_chaos_2017}, analogous to the result
obtained in Ref. \cite{del_campo_scrambling_2017}.

The above result connecting the quantum operation
\(\mathcal{U}_{\beta,t}\) to the two-point SFF makes one wonder if a direct
relation between the SFF and the regularized OTOC can be obtained,
since the \(\mathcal{U}_{\beta,t}\) originates in the choice of the
regularization for the thermal OTOC \cite{maldacena_bound_2016}. We
will now show that the regularized OTOC, averaged over \textit{global}
random unitaries is related to the \textit{four-point} SFF. Notice
that, unlike the bipartite averaging that we will focus on throughout this paper, this
relies on \textit{global} averages over the unitary group. The
necessity of performing global averages to connect with SFF subtly hints at the nonlocality (in both space and time) of
the SFF, see Refs. \cite{cotler_black_2017, cotler_chaos_2017} for a
detailed discussion. Let
\begin{equation}
F^{(A,B,C,D)}_{\beta}(t):= \operatorname{Tr}\left[ y A_{t} y B y C_{t} y D \right]
\end{equation}
with \(y = \rho_{\beta}^{1/4}\), then we have the following result.\\

\begin{restatable}{prop}{sffthermalotoc}
\label{prop:sff-thermal-otoc}
The regularized four-point OTOC averaged globally over Haar-random
unitaries is related to the four-point spectral form factor as,
\(\mathbb{E}_{A_{1},B_{1},A_{2} \in \mathcal{U}(\mathcal{H})} \left[
  F_{\beta}^{(A_{1},B_{1},A_{2},B_{2})}(t) \right] =
\mathcal{R}^{(H)}_{4}(\beta/4,t)/ \left( d^{3}
  \mathcal{Z}(\beta) \right)\), where \(B_{2} = A^{\dagger}_{2} B^{\dagger}_{1} A^{\dagger}_{1}\) and \(\mathcal{R}^{(H)}_{4}(\beta,t) := \left( \mathcal{Z}_{\beta}(t) \mathcal{Z}_{\beta}(t)^{*} \right)^{2}\) with \(\mathcal{Z}_{\beta}(t) = \operatorname{Tr}\left[ \exp \left[ (-\beta + it ) H\right]  \right]\), the analytically continued partition function.
\end{restatable}
Moreover, notice that this formula can be easily generalized to the case of different regularizations of the OTOC, for example, if we have \(2\)-point functions with \(\sqrt{\rho_{\beta}}\) inserted between them, then we can get the \(\mathcal{R}_{2}(\beta/2,t)\). In the most general case, if we have, \(2k\)-point thermally regulated OTOCs (which will have \(\rho_{\beta}^{1/2k}\) inserted between them), then, this will connect with \(\mathcal{R}_{2k}(\beta/2k,t)\).

\prlsection{Purity of the thermofield double} We are now ready to focus again on the local properties that are
quantified by the BROTOC. Let us consider a
bipartition of the original Hilbert space, \(\mathcal{H} \cong \mathcal{H}_{A} \otimes
\mathcal{H}_{B}\). Then the Choi matrix corresponding to the CP map
\(\mathcal{U}_{\beta,t}\) is a four-partite state, since
\(\rho_{\mathcal{U}_{\beta,t}} \in \mathcal{L}(\mathcal{H}_{A} \otimes
\mathcal{H}_{B}) \otimes \mathcal{L}(\mathcal{H}_{A'} \otimes
\mathcal{H}_{B'}) \), where the primed Hilbert spaces represent a
replica of the original Hilbert space. We can then compute the
\(2\)-norm squared of the reduced Choi matrix
\(\rho^{AA'}_{\mathcal{U}_{\beta,t}} \equiv \operatorname{Tr}_{BB'}\left[ \rho_{\mathcal{U}_{\beta,t}} \right]\) (or the purity if the matrix
was normalized; it is already positive semidefinite). Then, a key
lemma from Ref. \cite{zanardi_entanglement_2001-1} shows that \(\left\Vert \rho^{AA'}_{\mathcal{U}_{\beta,t}}
\right\Vert_{2}^{2} = \frac{1}{d^{2}} \operatorname{Tr}\left[
  \mathbb{S}_{AA'} \mathcal{U}_{\beta,t}^{\otimes 2} \left(
    \mathbb{S}_{AA'} \right) \right]\). Therefore, the (connected
component of the) regularized OTOC,
\begin{align}
  G^{(r)}_{\beta}(t) = \frac{d}{\mathcal{Z}(\beta)} \left\Vert \rho^{AA'}_{\mathcal{U}_{\beta,t}}
\right\Vert_{2}^{2}.
\end{align}
That is, it is proportional to the \(2\)-norm squared of the reduced
Choi matrix for the quantum operation \(\mathcal{U}_{\beta,t}\).

Now, let \(\mathcal{P}_{AA'}( | \psi \rangle_{ABA'B'}) :=  \left\Vert \operatorname{Tr}_{BB'}\left[ | \psi \rangle \langle  \psi |_{ABA'B'}
\right] \right\Vert_{2}^{2}\) be the purity of the thermofield double
across the \(AA'|BB'\) partition. Then, using the fact that the Choi matrix of
\(\mathcal{U}_{\beta,t}\) is proportional to the time-evolved thermofield double
state \(| \psi(\beta/2,t) \rangle \langle  \psi(\beta/2,t) | \), we have,
\begin{align}
  G^{(r)}_{\beta}(t) = \frac{\mathcal{Z}(\beta/2)^{2}}{d \mathcal{Z}(\beta)} \mathcal{P}_{AA'} \left( | \psi(\beta/2,t) \rangle_{ABA'B'} \right).
\end{align}
Finally, notice that Page scrambling of a quantum state
\cite{Sekino_2008, lashkari_towards_2013, hosur_chaos_2016} is defined as all subsystems containing less than
half the degrees of freedom being nearly maximally mixed. Since the
purity is minimal for maximally mixed states, the closer the value of
\(G^{(r)}_{\beta}(t)\) to the lower bound \(
\frac{\mathcal{Z}(\beta/2)^{2}}{d d_{A}^{2} \mathcal{Z}(\beta)}\), the
more information scrambling we have in the system's dynamics. That is, the connected component of
the BROTOC quantifies the degree of Page
scrambling in the time-evolving TDS. Furthermore, the connection to the purity of the thermofield double immediately
allows us to infer the following bounds (which also follow from the
connection to operator purity above),
\begin{align}
   \frac{\mathcal{Z}(\beta/2)^{2}}{d d_{A}^{2} \mathcal{Z}(\beta)} \leq G^{(r)}_{\beta}(t) \leq \frac{\mathcal{Z}(\beta/2)^{2}}{d \mathcal{Z}(\beta)},
\end{align}
where we have used the fact that the purity of a quantum state in
\(\mathcal{H}_{AA'}\) is bounded between \(\frac{1}{d_{A}^{2}}\) and \(1\).

\prlsection{Non-Hermitian evolution} The connected BROTOC has the form
\begin{align}
G^{(r)}_{\beta}(t) = \frac{1}{d \mathcal{Z}(\beta)} \left\langle \mathbb{S}_{AA'}, \mathcal{U}_{\beta,t}^{\otimes 2} \left( \mathbb{S}_{AA'} \right) \right\rangle,
\end{align}
which quantifies the autocorrelation function between the observable
\(\mathbb{S}_{AA'}\) and its evolved version
\(\mathcal{U}_{\beta,t}^{\otimes 2} \left( \mathbb{S}_{AA'}
\right)\). Now, recall that a non-Hermitian
Hamiltonian is usually defined to be of the form, \(H = H_{0} - i
\Gamma\), where \(H_{0}, \Gamma\) are Hermitian operators and we have
separated the Hermitian and non-Hermitian parts explicitly. Assume
that we are in the simple scenario where the Hermitian and
anti-Hermitian parts commute, namely, \(\left[ H_{0}, \Gamma \right]
=0\). Therefore, the time-evolution of an observable \(X\) under such
dynamics is given as \(X_{t} = e^{-\Gamma t} e^{i H_{0} t} X e^{-i
  H_{0} t} e^{-\Gamma t}\). For the connected BROTOC, if we identify \(\Gamma =
\beta H/(4t)\) at \(t>0\), then we can think of
\(\mathcal{U}_{\beta,t}\) as a simple \textit{non-Hermitian}
evolution (and the commutation assumption above is trivially
satisfied). In this case, the BROTOC quantifies the scrambling power of non-Hermitian dynamics. This identification opens up
the possibility of utilizing tools from the theory of dissipative
quantum chaos \cite{sa_spectral_2020-1, denisov_universal_2019,
  can_random_2019,PhysRevLett.123.234103,PhysRevLett.61.1899,PhysRevLett.123.254101,
  sa_complex_2020} such as complex spacing ratios\cite{sa_complex_2020}, to analyze directly the spectral correlations
encoded in the non-Hermitian dynamics \(\mathcal{U}_{\beta,t}\) as a means of distinguishing
integrable and chaotic dynamics. Furthermore, the ability to distinguish quantum
chaos from decoherence is a fascinating question with a long history
\cite{haake_quantum_2010, zurekDecoherenceChaosSecond1994}. Rewriting
the quantum operation \(\mathcal{U}_{\beta,t} = \mathcal{V}_{\beta}
\circ \mathcal{U}_{t}\) as a composition of a quantum operation  \(\mathcal{V}_{\beta}\) (which
signifies decoherence) and the unitary dynamics \(\mathcal{U}_{t}\) may allow for
disentangling the decoherence effects from the \textit{unitary scrambling}.

\subsection{Zero-temperature limit}
\label{sec:zero-temperature}
As we discussed above, the infinite-temperature limit of \(N_{\beta}(t)\)
is the operator entanglement of the unitary \(U_{t}\) and enjoys
several information-theoretic connections \cite{styliaris_information_2021}. What about the other limit,
namely, \(\beta \rightarrow \infty\)? Here, we show that in the
zero-temperature limit, the regularized OTOC probes the
operator purity of the ground state projector, depending on the
degeneracy of the ground state manifold. Let \(\Pi_{0}\) be the
groundstate projector, then, recall that, \(\lim\limits_{\beta
  \rightarrow \infty} \rho_{\beta} \rightarrow \Pi_{0}/g_{0}\), where \(g_{0}\) is the groundstate
degeneracy. That is, at zero temperature, the Gibbs state is
proportional to the projector onto the groundstate manifold. Moreover, since \(\Pi_{0}\) is a projector, we
have, \(\Pi_{0}^{2} = \Pi_{0}\). Therefore, the disconnected
correlator simplifies to,
\begin{align}
  F^{(d)}_{\beta \rightarrow \infty} = \frac{1}{g^{2}_{0}} \operatorname{Tr}\left[
  \Pi_{0} V^{\dagger} \Pi_{0} V \right] \operatorname{Tr}\left[
  \Pi_{0} W^{\dagger} \Pi_{0} W \right].
\end{align}
Similarly, for the
regularized part we have,
\begin{align}
  F^{(r)}_{\beta \rightarrow \infty}(t) = \frac{1}{g_{0}}
\operatorname{Tr}\left[ \Pi_{0} W_{t}^{\dagger} \Pi_{0} V^{\dagger}
  \Pi_{0} W_{t} \Pi_{0} V \right].
\end{align}
Now, let \(H = \sum\limits_{j}^{}
E_{j} \Pi_{j}\) be the spectral decomposition of the Hamiltonian,
then, the projectors \(\{ \Pi_{j} \}_{j}\) are orthonormal (but not
necessarily rank-\(1\)). Plugging in \(U_{t} = \sum\limits_{j}^{} \exp
\left[ -i E_{j}t \right] \Pi_{j}\), we get,
\begin{align}
  F^{(r)}_{\beta
  \rightarrow \infty}(t) = \frac{1}{g_{0}}
\operatorname{Tr}\left[ \Pi_{0} W^{\dagger} \Pi_{0} V^{\dagger}
  \Pi_{0} W \Pi_{0} V \right].
\end{align}
Now, if the groundstate is
nondegenerate, then, we have, \(\Pi_{0} = | \psi_{\mathrm{gs}} \rangle
\langle  \psi_{\mathrm{gs}} | \), where \(| \psi_{\mathrm{gs}}
\rangle\) is the ground state wavefunction and \(g_{0}=1\). Then, a simple calculation
shows that, for this case,
\begin{align}
  F^{(d)}_{\beta} =
\left| \langle \psi_{\mathrm{gs}} | V |  \psi_{\mathrm{gs}} \rangle
\right|^{2} \left| \langle \psi_{\mathrm{gs}} | W |
  \psi_{\mathrm{gs}} \rangle \right|^{2} = F^{(r)}_{\beta}
\end{align}
and therefore, their difference vanishes. In fact, notice that, the
four-point correlator has now reduced to a product of \(1\)-point
correlators. In summary, at
zero temperature, for nondegenerate Hamiltonians, the correlator \(F^{(d)}_{\beta} -
F^{(r)}_{\beta}(t)\) vanishes, and so does the regularized bipartite
OTOC \(N_{\beta \rightarrow \infty}(t)\).

We now perform the bipartite averaging for the zero-temperature case,
without the assumption of nondegeneracy. The following result establishes that if the ground state is degenerate, then, both the
disconnected and connected components of the regularized bipartite
OTOC probe the entanglement in the ground state projector. Moreover,
for the nondegenerate case, both terms are proportional to the square
of the purity of the ground state and can be utilized to detect quantum
phase transitions \cite{sachdev_quantum_2011}. The ability of
groundstate OTOCs to detect quantum
phase transitions was explored in
Ref. \cite{PhysRevLett.121.016801}. Establishing a possible connection
to finite-temperature phase transitions is an interesting question for future investigations.\\

\begin{prop}
\label{prop:zero-temperature}
The disconnected and connected components of the bipartite averaged
OTOC at zero temperature are,
\begin{align}
&G^{(d)}_{\beta \rightarrow \infty} = \frac{1}{dg_{0}^{2}}
                \mathcal{P}_{A}(\Pi_{0}) \mathcal{P}_{B}(\Pi_{0}),
                \text{ and }\nonumber \\
&G^{(r)}_{\beta \rightarrow \infty} = \frac{1}{dg_{0}} \operatorname{Tr}\left[ \mathbb{S}_{AA'} \Pi_{0}^{\otimes 2} \mathbb{S}_{AA'} \Pi_{0}^{\otimes 2}\right].
\end{align}
\end{prop}
Note that both quantities becomes \textit{time-independent} and the
convergence to the groundstate is \textit{exponential} in \(\beta\),
given the Gibbs weights. Finally, we note that for a pure quantum
state \(\Pi = | \psi \rangle \langle  \psi | \), we have,
\(\operatorname{Tr}\left[ \mathbb{S}_{AA'} \Pi^{\otimes 2}
  \mathbb{S}_{AA'} \Pi^{\otimes 2} \right] = \left\Vert \rho_{A}
\right\Vert_{2}^{4}\), where \(\rho_{A} \equiv
\operatorname{Tr}_{B}\left[ | \psi \rangle \langle  \psi |  \right]\)
\cite{zanardi_entanglement_2001-1}. That is, the \textit{operator}
purity term reduces to the \textit{state} purity squared. Therefore,
the connected (and disconnected) BROTOC at zero temperature, for a nondegenerate Hamiltonian probes its groundstate purity.

\subsection{Long-time limit and eigenstate entanglement}
\label{sec:longtime-limit}
The equilibration value of correlation functions has long been studied
as a probe to thermalization and chaos \cite{dalessio_quantum_2016,
  borgonovi_quantum_2016}. Although, for finite-dimensional quantum
systems, correlation functions typically do not converge to a limit
for \(t \rightarrow \infty\). Instead, after a transient initial
period, they oscillate around some equilibrium value
\cite{reimann_foundation_2008, linden_quantum_2009,
  campos_venuti_exact_2011, alhambra_time_2020}, which can be
extracted via long-time averaging (also known as infinite-time averaging), defined as, \(\overline{A(t)} :=
\lim\limits_{T \rightarrow \infty} \frac{1}{T} \int\limits_{0}^{T}
A(\tau) d \tau\). In Refs. \cite{styliaris_information_2021,
  anand_quantum_2020, garcia-mata_chaos_2018, fortes_gauging_2019, huang_finite-size_2019},
the equilibration value of the OTOC (or the averaged OTOC) was used to
distinguish integrable versus chaotic quantum systems. Here, we
discuss how the long-time average of the BROTOC
can also reveal the degree of integrability for Hamiltonian quantum
systems, and discuss the \(\beta\)-dependence.

A key assumption on the energy spectrum  that we will use in this section is the so-called
no-resonance condition (NRC) or nondegenerate energy gaps
condition \cite{reimann_foundation_2008, shortEquilibrationQuantumSystems2011}. Simply put, both the energy levels and the energy gaps
between these levels is nondegenerate. More formally, consider the
spectral decomposition of the Hamiltonian, \(H =
\sum_{j=1}^{d} E_{j} | \phi_{j} \rangle \langle  \phi_{j} |\). Then,
\(H\) obeys NRC if \(E_{l} + E_{k} = E_{n} + E_{m} \iff l=n,k=m \text{
or } l=m,k=n ~~\forall j,k,l,m\). The NRC condition is satisfied by
\textit{generic} quantum systems and in particular, chaotic quantum
systems satisfy such a condition either exactly or to a close
approximation. Let us denote by \(\rho^{\chi}_{j} := \operatorname{Tr}_{\overline{\chi}}\left[ | \phi_{j} \rangle \langle  \phi_{j} |  \right], \chi = \{ A,B \}\) the reduced density matrix corresponding to the \(j\)-th Hamiltonian eigenstate. Moreover, we introduce a Gram matrix corresponding to the inner product between the reduced states, \(R^{(\chi)}_{jk} := \left\langle \rho_{j},\rho_{k} \right\rangle\) with \(\chi = \{ A,B \}\) and \(\left\langle \cdot, \cdot \right\rangle\) the Hilbert-Schmidt inner product. Then, we have the following result. \\
\begin{restatable}{prop}{longtimeaverage}
\label{prop:longtimeavg}
\begin{align}
  \overline{G_{\beta}^{(r)}(t)}& = \frac{1}{d\mathcal{Z}(\beta)} \left[ \sum\limits_{j,k=1}^{d} \exp \left[ -\beta (E_{j} + E_{k})/2 \right] \right. \nonumber\\
  &\left. \left( \left| R_{jk}^{A} \right|^{2} + \left| R_{jk}^{B} \right|^{2} - \delta_{jk} \left| R_{jk}^{A} \right|^{2} \right) \right].
\end{align}
\end{restatable}

This result generalizes to finite-temperature the Proposition 4 obtained in \cite{styliaris_information_2021} and therefore at \(\beta=0\), reduces to the form described there. We can rescale the reduced states as \(\sigma_{j}^{\chi}:= \exp \left[
  -\beta E_{j}/4 \right] \rho_{j}^{\chi}\), which generates a rescaled
Gram matrix \(\widetilde{R}^{(\chi)}_{jk} := \left\langle
  \sigma_{j}^{\chi}, \sigma_{k}^{\chi} \right\rangle = \exp \left[
  -\beta \left( E_{j} + E_{k} \right)/4 \right] \left\langle
  \rho_{j}^{\chi}, \rho_{k}^{\chi} \right\rangle\). Therefore, we can
rewrite the time-average as,
\begin{align}
\overline{G_{\beta}^{(r)}(t)} = \frac{1}{d \mathcal{Z}(\beta)} \sum\limits_{\chi \in \{ A,B \}}^{} \left( \left\Vert \widetilde{R}^{(\chi)} \right\Vert_{2}^{2} - \frac{1}{2} \left\Vert \widetilde{R}_{D}^{(\chi)} \right\Vert_{2}^{2} \right),
\end{align}
with \([ \widetilde{R}_{D}^{(\chi)} ]_{jk} = [
\widetilde{R}_{D}^{(\chi)} ]_{jk}  \delta_{jk}\).

Similarly, for the disconnected correlator, we have,
\begin{align}
&  G^{(d)}_{\beta} = \frac{1}{d} \left\Vert \operatorname{Tr}_{A}\left[ \sqrt{\rho_{\beta}} \right] \right\Vert_{2}^{2} \left\Vert \operatorname{Tr}_{B}\left[ \sqrt{\rho_{\beta}}  \right] \right\Vert_{2}^{2} \nonumber\\
  &= \frac{1}{d} \left\Vert \sum\limits_{j=1}^{d} \frac{\exp \left[ -\beta E_{j}/2 \right] }{\sqrt{\mathcal{Z}(\beta)}} \rho_{j}^{B} \right\Vert_{2}^{2} \left\Vert \sum\limits_{k=1}^{d} \frac{\exp \left[ -\beta E_{k}/2 \right] }{\sqrt{\mathcal{Z}(\beta)}} \rho_{k}^{A} \right\Vert_{2}^{2} \nonumber\\
&= \frac{\mathcal{Z}(\beta/2)^{4}}{d \mathcal{Z}(\beta)^{2}} \left\Vert \sum\limits_{j=1}^{d} p_{j}(\beta/2) \rho_{j}^{B} \right\Vert_{2}^{2} \left\Vert \sum\limits_{k=1}^{d} p_{k}(\beta/2) \rho_{k}^{A} \right\Vert_{2}^{2},
\end{align}
where \(p_{j}(\beta):= \exp \left[ -\beta E_{j} \right]/\mathcal{Z}(\beta)\) is the Gibbs probability associated to the energy level \(j\) at inverse temperature \(\beta\).

\textit{Maximally-entangled models}.--- \autoref{prop:longtimeavg} allows us to connect the equilibration
value of the regularized OTOC with the entanglement in the Hamiltonian
eigenstates. As a concrete example, we
evaluate this equilibration value for a symmetric bipartition, that
is, \(d_{A} = d_{B} = \sqrt{d}\)  and a Hamiltonian whose eigenstates
are maximally entangled, that is, \(\{ | \phi_{k} \rangle
\}_{k=1}^{d}\) are maximally entangled across the \(A|B\) partition. We term this Hamiltonian a ``maximally entangled Hamiltonian'' for brevity. For such a Hamiltonian, we have, \(\rho_{k}^{A} = \mathbb{I}/\sqrt{d} = \rho_{k}^{B} ~~\forall k\) and therefore, \(R_{k,l}^{A} = \operatorname{Tr}\left[ \rho_{k}^{A} \rho_{l}^{A} \right] = \frac{1}{d} \operatorname{Tr}\left[ I_{\sqrt{d}} \right] = \frac{1}{\sqrt{d}} = R_{kl}^{B} ~~\forall k,l\). Then, we have,
\begin{align}
\label{prop:brotoc-maxent}
\overline{G^{(r)}_{\beta}(t)} \left. \right|_{\text{ME}} = \frac{1}{d^{2}} \left( \frac{2 \mathcal{Z}(\beta/2)^{2}}{\mathcal{Z}(\beta)} -1 \right).
\end{align}
Notice that this equilibration value is close to the lower bound (for a symmetric bipartition): \(\mathcal{Z}(\beta/2)^{2}/d^{2}\mathcal{Z}(\beta) \leq G^{(r)}_{\beta}(t)\).

Similarly, for the disconnected correlator, one can show that,
\begin{align}
G^{(d)}_{\beta} \left. \right|_{\text{ME}} = \frac{\mathcal{Z}(\beta/2)^{4}}{d^{2} \mathcal{Z}(\beta)^{2}}.
\end{align}
Putting everything together, we have, the equilibration value of the BROTOC for a maximally-entangled Hamiltonian is,
\begin{align}
\label{eq:longtime-avg-maximal}
\overline{N_{\beta}(t)} \left. \right|_{\text{ME}} = \frac{1}{d^{2}} \left( \frac{\mathcal{Z}(\beta/2)^{2}}{\mathcal{Z}(\beta)} - 1 \right)^{2}. 
\end{align}
Notice that at \(\beta=0\), \(\mathcal{Z}(\beta) =
\operatorname{Tr}\left[ I \right] = d =
\mathcal{Z}(\beta/2)\). Therefore, the above evaluates to
\(\overline{N_{\beta =0}(t)} \left. \right|_{\text{ME}} = \left( 1 -
  \frac{1}{d} \right)^{2} = \overline{G}^{\mathrm{NRC}}_{\text{ME}}\)
for \(\beta = 0\) as in Ref. \cite{styliaris_information_2021}, which
shows that the equilibration value is nearly maximal; which for a
symmetric bipartition is equal to \(1 - 1/d\). For quantum chaotic
systems, random matrix theory predicts that the spectral and
eigenstate properties of the Hamiltonian resemble those of the
Gaussian random matrix ensembles (depending on the universality class) \cite{guhr_random-matrix_1998,mehta_random_2004}, which typically have nearly maximally entangled eigenstates. Therefore, one can expect the equilibration value to be close to \cref{eq:longtime-avg-maximal}.

We outline here a qualitative argument to understand the decrease of $\overline{G^{(r)}_\beta(t) }$ with $\beta$ as will become evident in the section on Numerical Simulations. In fact, this monotonicity is an \textit{entropic} effect due to the fact that, by increasing $\beta$, less and less states contribute to the sum in \autoref{prop:longtimeavg}, the general formula for all Hamiltonians that satify NRC. We now make a quantitative argument: let \({\mathbf{p}}(\beta/2)\) be a probability vector whose components are \(p_i=\frac{e^{-\beta E_i/2}}{Z(\beta/2)}\). Then, \(\|{\mathbf{p}}(\beta/2)\|^2=Z(\beta)Z(\beta/2)^{-2}\) and we can reexpress the BROTOC as,
\begin{align}
\overline{G^{(r)}_\beta(t) }= \frac{ \langle{ \mathbf{p}}(\beta/2), \hat{C}{\mathbf{p}}(\beta/2)\rangle}{d\|{\mathbf{p}}(\beta/2)\|^2}, \label{G_r}
\end{align}
where \(C_{ij}:= |R_{ij}^A|^2+|R_{ij}^B|^2-\delta_{ij}  |R_{ij}^A|^2\).

The denominator of \cref{G_r} is proportional to the purity of ${\mathbf{p}}(\beta/2)$ and it is therefore monotonically increasing with $\beta$
($d^{-1}$ at $\beta=0$, and $1$ at $\beta=\infty$.) On the other hand, the numerator  of \cref{G_r} can change from $O(d_A^{-2})$ at $\beta=0$ to $O(1)$ for local models ($O(d_A^{-2})$ for non-local ones). This change is always dominated by the purity increase in the denominator. For example in the maximally entangled case ($d_A=d_B=\sqrt{d})$) one has $C_{ij}= d^{-1}(2-\delta_{ij})$ and therefore one can rewrite \cref{prop:brotoc-maxent} as
\begin{align}
\overline{G^{(r)}_\beta(t) }|_{\mathrm{ME}} &= \frac{2-\|{\mathbf{p}}(\beta/2)\|^2}{d^2\|{\mathbf{p}}(\beta/2)\|^2} \nonumber\\
&=\frac{1}{d^2}\frac{1+S_ \mathrm{lin}(\beta/2)}{1-S_ \mathrm{lin}(\beta/2)}
, \label{G_rr}
\end{align}
where $S_ \mathrm{lin}(\beta):=1-\|{ \mathbf{p}}(\beta)\|^2$ is the \textit{linear entropy}  of $\rho(\beta)$. This function is clearly monotonically decreasing with $\beta$ and shows, once again, that the increase of of the time-averaged connected OTOC with temperature is an \textit{entropic} effect.

\prlsection{Nearly maximally-entangled models} We will now show that, if the Hamiltonian eigenstates are highly entangled then it implies a bound on the equilibration value that is close to the maximally entangled case. Recall that a quantum state is called ``Page scrambled'' \cite{Sekino_2008, lashkari_towards_2013, hosur_chaos_2016} if any arbitrary
subsystem that consists of up to half of the state's degrees of
freedom are nearly maximally mixed. In the following proposition, we
assume Page scrambling of all Hamiltonian eigenstates across a
symmetric bipartition \(d_{A} = d_{B} = \sqrt{d}\) and show that the
equilibration value is close to that of highly entangled models. Let \(\mathcal{P}(|
\psi_{AB} \rangle)\) denote the purity of the reduced state of \(|
\psi_{AB} \rangle\) across the bipartition \(A|B\) and
\(\mathcal{P}_{\min} = \min \{ \frac{1}{d_{A}}, \frac{1}{d_{B}} \}\)
be the minimum purity of a quantum state across the \(A|B\)
bipartition. Recall that a pure state \(| \psi_{AB} \rangle\) is maximally
entangled across \(A|B\) \(\iff \mathcal{P}(| \psi_{AB} \rangle) =
\mathcal{P}_{\min}\). Then, the deviations from the maximally
entangled value are bounded as follows.\\

\begin{restatable}{prop}{eigenstateentanglement}
  \label{prop:eigenstate-entanglement}
For a symmetric
bipartition of the Hilbert space, \(d_{A} = d_{B} = \sqrt{d}\) if \(\mathcal{P}(| \psi_{AB} \rangle)-\mathcal{P}_{\min} \leq
\epsilon\) holds for all eigenstates, then for systems satisfying NRC,
the equilibration value is bounded away from the maximally entangled
case as follows,
\(\left| \overline{G^{(r)}_{\beta}(t)} ~|_{\mathrm{ME}} - \overline{G^{(r)}_{\beta}(t)} ~|_\mathrm{NRC} \right|  \leq \frac{\mathcal{Z}(\beta/2)^{2}}{d \mathcal{Z}(\beta)}  \left( \frac{6 \epsilon}{\sqrt{d}} + 3\epsilon^{2} \right) \).
\end{restatable}

\begin{figure}[!th]
     \centering
 \includegraphics[width=0.5\textwidth]{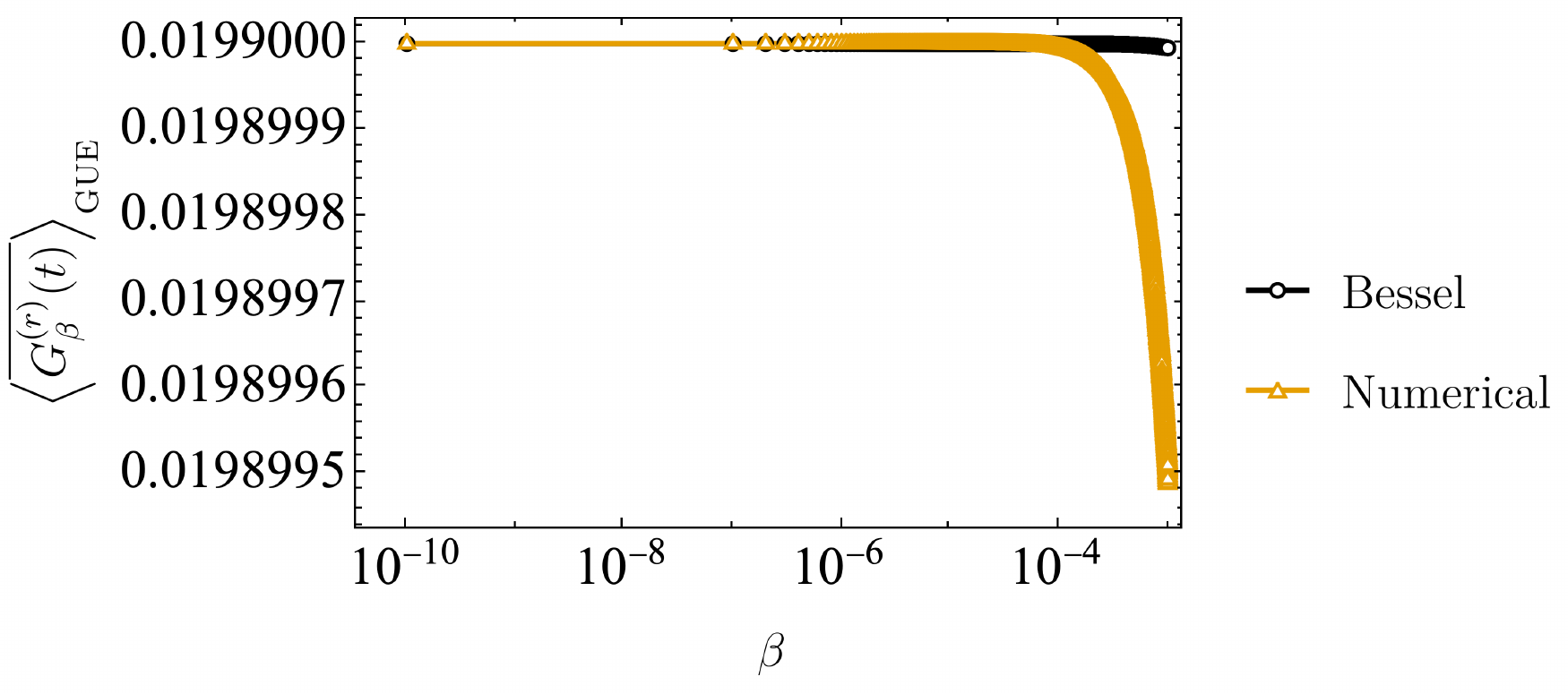}
\caption{A log-log plot of the equilibration value (long-time average)
  of the \(\overline{G}_{\beta}^{(r)}(t)\) for the GUE Hamiltonian at
  \(d=100\) for \(10^{-10}  \leq \beta \leq 10^{-3}\) comparing the numerical
  estimate to the Bessel function form above.}
    \label{fig:gue-bessel}
\end{figure}

\textit{Unregularized vs regularized OTOC}.--- We highlight a key
difference between the bipartite regularized versus unregularized
OTOCs. As we will note, for nearly maximally entangled models, the
\(G_{\beta}(t)\) is nearly \(\beta\)-independent, while the
\(G_{\beta}^{(r)}(t)\) shows a clear \(\beta\)-dependence as we have seen above. The proof relies on using an operator Schmidt decomposition for
the unitary \(U_{t}\), see the Appendix for more details. We obtain that,
\begin{align}
  G_{\beta}(U) = 1 - F_{\beta}(U) = 1 - \frac{1}{d_{A}^{2}},
\end{align}
and is independent of \(\beta\). Contrast this, with the equilibration value for nearly maximally
entangled Hamiltonians \cref{eq:longtime-avg-maximal}, as computed above. Let us consider
Hamiltonians from the Gaussian Unitary Ensemble (GUE) as an
example. The eigenstates of these are known to have near maximal
entanglement and therefore, we can approximate the
\(\overline{N_{\beta}(t)} \approx \frac{1}{d^{2}} \left(
  \frac{\mathcal{Z}(\beta/2)^{2}}{\mathcal{Z}(\beta)} - 1
\right)^{2}\). Moreover, for large-\(d\), the partition function after ensemble
averaging can be expressed as \cite{mehta_random_2004, vijay_finite-temperature_2018}, \(\left\langle
  \mathcal{Z}(\beta) \right\rangle_{\mathrm{GUE}} = \frac{d I_{1}(2
  \beta)}{\beta}\), where \(I_{1}(\beta)\) is the modified Bessel
function of the first kind. Therefore, the ensemble averaged
equilibration value of \(N_{\beta}(t)\) for the GUE is \footnote{We
  have implicitly assumed here that the ensemble averaging and the
  large-\(d\) limits commute, see \cite{cotler_black_2017, xu_thermofield_2020} for a discussion.}
\begin{align}
\left\langle \overline{N_{\beta}(t)} \right\rangle_{\mathrm{GUE}} \approx \left[ \frac{4 I_{1}(\beta)^{2}}{\beta I_{1}(2 \beta)} - \frac{1}{d}	\right]^{2}. 
\end{align}
Notice that, \(I_{1}(2 \beta)/\beta =
\sum\limits_{n=0}^{\infty} \frac{\beta^{2n}}{\left( n! \right)^{2}
  (n+1)}\). Therefore, we can extract from this, both the low- and
high-temperature estimates. In \cref{fig:gue-bessel} we plot the
Bessel function form along with the numerical estimate of the
long-time average of the connected BROTOC for the GUE, obtained by averaging (numerically generated) GUE Hamiltonians for \(d=100\).

\prlsection{NRC-product states (NRC-PS)} We introduce a Hamiltonian model that has a generic spectrum, namely, one that satisfies NRC but with all eigenstates as product states (for example, the computational basis states), that we call ``NRC-PS''. The Hamiltonian can be expressed as,
\begin{align}
H_{\mathrm{NRC\text{-}PS}}:=
  \sum\limits_{j,k=1}^{d_{A},d_{B}} E_{j,k} | \phi_{j}^{(A)} \rangle
  \langle  \phi_{j}^{(A)} | \otimes | \phi_{k}^{(B)} \rangle
  \langle  \phi_{k}^{(B)} |,
\end{align}  
where the spectrum \(\{ E_{j,k} \}_{j,k}\) satisfies NRC\footnote{Note that any Hamiltonian that satisfies NRC cannot be noninteracting, i.e., cannot be of the form \(H = H_{A} \otimes \mathbb{I}_{B} + \mathbb{I}_{A} \otimes H_{B}\) since such a Hamiltonian would, by construction, violate NRC. As an example, consider product eigenstates of the form, \(\{ | \phi_{j}^{(A)} \rangle \otimes | \chi_{k}^{(B)} \rangle \}_{j,k}\) then it is easy to find pairs of eigenstates for which the energy gaps are equal \cite{shortEquilibrationQuantumSystems2011}. However,
the converse is not true, namely, there exist interacting Hamiltonians, i.e., of the form \(H \neq H_{A} \otimes \mathbb{I}_{B} + \mathbb{I}_{A} \otimes H_{B}\) that have product eigenstates.}; for example, consider the spectrum of
  a Hamiltonian from a Gaussian Unitary Ensemble (GUE). The reason to introduce such a model is twofold: first, it allows us to \textit{disentangle} the spectral and eigenstate contributions to the equilibration value \(\overline{G^{(r)}_{\beta}(t)}\) since, it has the spectrum of a ``chaotic'' model and the eigenstate properties of a ``free'' model. Second, as we show now, this model is analytically tractable. The key reason for this is that the NRC-PS model has an extensive number of conserved quantities, \(d = 2^{L}\) of them in fact. A local operator of the form, \(A_{j} = | \phi_{j}^{A} \rangle \langle  \phi_{j}^{A} | \otimes \mathbb{I}\) commutes with the Hamiltonian above, \(\left[ H,A_{j} \right] =0 ~~\forall j \in \{ 1,2, \cdots, d_{A} \}\). Similarly, operators of the form, \( B_{k} = \mathbb{I} \otimes | \phi_{k}^{B} \rangle \langle  \phi_{k}^{B} |\) also commute with the Hamiltonian,  \(\left[ H,B_{k} \right] =0 ~~\forall k \in \{ 1,2, \cdots, d_{B} \}\). Therefore, the Hamiltonian has \(d = d_{A} d_{B}\) number of local conserved quantities. In this sense, this is an integrable model, notice however, that its spectrum is intentionally chosen to satisfy NRC.

The presence of conserved quantities enable an exact calculation for the equilibration value of the BROTOC in this model. A detailed proof of this can be found in the Appendix.
\begin{align}
\overline{G^{(r)}_\beta(t) }|_{\mathrm{NRC-PS}}=\frac{1}{d}\left( \frac{ \|{ \mathbf{p}}^A(\beta/2)\|^2+\| {\mathbf{p}}^B(\beta/2)\|^2}{\| {\mathbf{p}}(\beta/2)\|^2}-1 \right) 
\label{G_NRC-PS}
\end{align}
where the probability vector ${ \mathbf{p}}(\beta)$ is as in the above and ${ \mathbf{p}}^{A/B}(\beta)$ are its marginals e.g.,
${ \mathbf{p}}_j^{A}(\beta)=\sum_{k=1}^{d_B}{ \mathbf{p}}_{jk}(\beta)=\frac{1}{Z(\beta)} \sum_{k=1}^{d_B} e^{-\beta E_{jk}}.$

From this  Equation one finds immediately $\overline{G^{(r)}_{\beta = 0}(t) }|_{\mathrm{NRC-PS}}= 1/d_A +1/d_B -1/d > 1/d$ at infinite temperature and, $\overline{G^{(r)}_{\beta = \infty}(t) }|_{\mathrm{NRC-PS}}=1/d$ at zero temperature. For a symmetric bipartition $d_A=d_B=\sqrt{d}$, the former simplifies to $2/\sqrt{d} -1/d =O(1/\sqrt{d})$. And, as we will see in the next section, the numerical data obtained from finite-size scaling in \cref{table-parametervalues-vs-beta,table-parametervalues-appendix} is consistent with this as \(\overline{G^{(r)}_{\beta}(t)} \approx \frac{2}{\sqrt{d}}\).

\section{Numerical simulations}
\label{sec:numerics}

In this section we study numerically various dynamical features of the
BROTOC. In particular, we vary the \textit{degree}
of integrability of Hamiltonian systems and quantify it's effect on
the equilibration value \(\overline{G^{(r)}_{\beta}(t)}\). At
\(\beta=0\), this is equal to one minus the operator entanglement of
the dynamical unitary, whose equilibration value was used to
distinguish various integrable and chaotic models in
Ref. \cite{styliaris_information_2021}, see also Refs. \cite{zhou_operator_2017, garcia-mata_chaos_2018, fortes_gauging_2019, fortes_signatures_2020} for distinguishing integrable and chaotic models via time-averages of the OTOC or operator entanglement. We also refer the reader to Ref. \cite{PhysRevB.104.104306}, where bounds on decay of OTOCs in time were obtained using the scaling of the time-averaged OTOC. Here, we perform more
extensive numerical studies, consider more generally the \(\beta\)-dependence of this quantity, and focus
on the following Hamiltonian models of interest:\\

\begin{enumerate}

  \item Integrable model: The transverse-field Ising model (TFIM) with the Hamiltonian, \(H_{\mathrm{TFIM}}= -
  \sum\limits_{j=1}^{L-1} \sigma_{j}^{z} \sigma_{j+1}^{z} - g \sum\limits_{j=1}^{L}
    \sigma_{j}^{x} - h \sum\limits_{j=1}^{L} \sigma_{j}^{z},\) as a paradigmatic
  quantum spin-chain model. Here, the
  \(\sigma_{j}^{\alpha}, \alpha \in \{ x,y,z \}\) are the Pauli
  matrices. For the TFIM, \(g,h\) denotes the strength of the
  transverse field and the local field, respectively. The TFIM
  Hamiltonian is integrable for either \(h=0\) or \(g=0\) and nonintegrable when both
  \(g,h\) are nonzero. We consider as the integrable point,
  \(g=1,h=0\) and the nonintegrable point \(g=-1.05,h=0.5\). At the
  integrable point, this model can be mapped onto free fermions via
  the Jordan-Wigner transformation and is ``highly integrable'' in
  this sense. At the nonintegrable point, the model is quantum
  chaotic, in the sense of random matrix spectral statistics \cite{PhysRevLett.106.050405,PhysRevLett.111.127205} and
  volume-law entanglement of eigenstates \cite{wolfAreaLawsQuantum2008}.\\

\item Localized models: We study Anderson and many-body localization (MBL) with
the Hamiltonian, \(H_{\mathrm{MBL}}= -
  \sum\limits_{j=1}^{L-1} \sigma_{j}^{z} \sigma_{j+1}^{z} - \sum\limits_{j=1}^{L}
  g_{j}  \sigma_{j}^{x} - h \sum\limits_{j=1}^{L} \sigma_{j}^{z},\)
  where we draw from the uniform distribution, each \(g_{j} \in
  \left[ -\eta,\eta \right] \). In the absence of the longitudinal field,
  i.e., \(h=0\), and for nonzero disorder, this (disordered) free fermion model is Anderson
  localized. In the presence of the longitudinal field, the fermions
  are interacting and at sufficiently strong disoder, the model is
  many-body localized (MBL). As is well-known, MBL escapes
  thermalization by emergent integrability \cite{nandkishoreManyBodyLocalizationThermalization2015, dalessio_quantum_2016}. We refer the reader to Ref. \cite{chen_out--time-order_2017} for a discussion of the long-time values of the \textit{unregularized} OTOC in localized phases. In our numerical simulations, we focus on \(\eta = 10\) for the disorder strength and \(h=0.1\) for the MBL case. We average each instance of the disordered model over \(\floor*{200/L}\) independent realizations. In each case, the error bars are too small to plot alongside the data points.\\

\item NRC-product states (NRC-PS): As introduced before this model
  allows us to separate the \textit{spectral} and \textit{eigenstate}
  contributions to the BROTOC's equilibration  value. We choose a ``chaotic'' spectrum (in the sense that it corresponds to a  GUE Hamiltonian and hence is an instance of a model that obeys Wigner-Dyson statistics), while having the eigenstates of a \textit{noninteracting} model, that is, simple product states. To study the NRC-PS model numerically, we generate a random matrix from the Gaussian Unitary Ensemble (GUE) and use its spectrum, while keeping product eigenstates. We average this numerically over \(\floor*{200/L}\) independent realizations. This yields numerical results consistent with the analytical expression obtained from \cref{G_NRC-PS}.

\item Random matrices: As a benchmark for a ``maximally chaotic'' model, we consider
  Hamiltonians drawn from the GUE. For
  Hamiltonian systems, the seminal works of Berry and Tabor
  \cite{berry1977level} and that of Bohigas, Giannoni, and Schmit
  \cite{bohigas_characterization_1984} establishes that Poisson
  level-statistics is a characteristic feature of integrable, while for thermalizing systems,
  Wigner-Dyson statistics are the norm
  \cite{rigol_thermalization_2008,
    nandkishoreManyBodyLocalizationThermalization2015}. Furthermore,
  the eigenstates of GUE are nearly maximally entangled and will
  provide an analytically tractable example of a highly chaotic model. For our numerical simulations, we generate random matrices and average over \(\floor*{200/L}\) independent realizations. The error bars are too small to plot alongside the data points in this case as well. The ``ME'' in the plots corresponds to a maximally entangled model for which we use the analytical expressions from \cref{prop:brotoc-maxent}. For this we generate the spectrum from the GUE and average over \(\floor*{200/L}\) independent realizations.\\

\end{enumerate}

Throughout this section, to evaluate the time-averages \(\overline{G^{(r)}_{\beta}(t)}\) numerically, we use two different methods. First, for the Anderson and TFIM integrable model, since it does not satisfy NRC (the Hamiltonian has symmetries), we perform exact time evolution. We do this for a time interval of \(t \in [10, 10^3]\) with a \(10^6\) time steps in between. This is fixed for all the system sizes \(L\) and inverse temperatures \(\beta\). All models except these two satisfy NRC (also verified numerically) and so we compute \(\overline{G^{(r)}_{\beta}(t)}\) using the analytical expression in \autoref{prop:longtimeavg}. To do this, we perform exact diagonalization of the full Hamiltonian for this and compute the reduced states. At large \(\beta\), it is easy to show that one only needs the ground state along with a few excited states to estimate the time-average in \autoref{prop:longtimeavg}. Therefore, for \(L=13,14\) and \(\beta = 1\), we only extract the lowest $20$ Hamiltonian eigenstates.

The first numerical result focusses on \(L=6\) qubits and we study the variation of \(\overline{G^{(r)}_{\beta}(t)}\) as a function of \(\beta\). In
\cref{fig:longtimeavg-vs-beta}, we notice the following
\textit{universal features} of \(\overline{G^{(r)}_{\beta}(t)}\) as a
function of \(\beta\): the equilibration value is very slowly decaying
as \(\beta\) varies from zero to \(O(1)\). Around \(\beta = O(1)\), the equilibration value quickly decays to the asymptotic
value. Using, \autoref{prop:zero-temperature}, we note that the
asymptotic value \(\beta \rightarrow \infty\) is proportional to the operator purity
for the ground state projector. With these universal features at hand,
we systematically study the equilibration value
\(\overline{G^{(r)}_{\beta}(t)}\) for three representative choices of \(\beta\): \(\beta=0,
\beta = 1\) and \(\beta \rightarrow \infty\). We numerically study
their scaling as a function of the system size for a symmetric
bipartition of the lattice, \(\floor*{L/2}:\ceil*{L/2}\). For the case where \(L\) is not even, the numerical results are very similar for either choice of bipartition, \(\floor*{L/2}:\ceil*{L/2}\) or  \(\ceil*{L/2}:\floor*{L/2}\), and therefore, we choose the former throughout. We also label as ``logplot''
a plot with logarithmic scale on the \(y\)-axis and ``loglogplot''
those with logarithmic scale on both \(x\)- and \(y\)-axes.

The results for \(\beta = 0\) are discussed in
\cref{fig:longtimeavg-beta-zero}. We notice that the scaling
w.r.t. the system size is
effectively divided into two classes: the quantum chaotic models,
namely, the nonintegrable TFIM and the GUE. And, the second class is
all the others, namely, the free fermions, the Anderson and MBL, and
the NRC-PS. These two classes are primarily distinguished by their
eigenstate entanglement, namely, the scaling of the entanglement
across the entire spectrum. This, perhaps, comes as no surprise since
the infinite-temperature OTOC by construction probes the entanglement
across all eigenstates.

At  \(\beta = 1\), from \cref{fig:longtimeavg-beta-one}, we notice
that the MBL and quantum chaotic models have merged, while having a
distinct scaling from the other integrable models and the GUE. Recall
that at \(\beta=0\), Theorem 6 of
Ref. \cite{styliaris_information_2021} establishes a hierarchy between
the equilibration values of various estimates for the equlibration
value. However, for the
regularized OTOC, this result does not necessarily hold away from the
\(\beta=0\) case since now we have extra \(H\)-dependent terms in the
NRC estimate; see \autoref{prop:zero-temperature}.

And finally, the scaling for \(\beta = \infty\) can be understood
using \autoref{prop:zero-temperature}. For the nondegenerate
Hamiltonians, this simply probes the ground state entanglement. We notice that all curves coalesce into two groups, one for the integrable/localized models and the second for the GUE and ME models, respectively. While these models vary in their degree of integrability, their ground states (apart from the GUE/ME) all follow an area law \cite{wolfAreaLawsQuantum2008} and hence obey a
different decay rate with \(L\) from the GUE. Note that the GUE ground state is a Haar random state and therefore, should scale as
\(\overline{G_{\beta \rightarrow \infty}^{(r)}(t)} \sim
\frac{1}{d^{2}}\), which is consistent with the finite-size scaling results.

\begin{table}[!h]
  \centering
  \begin{tabular}{||c |c| c | c||}
 \hline
 Model & \(\beta = 0\) & \(\beta = 1\) & \(\beta = \infty \)\\ [0.5ex] 
 \hline\hline
TFIM integrable & 0.507672 & 0.687979 & 1.01858 \\ 
 \hline
NRC-PS & 0.495827 & 0.7218 & 1.00 \\
 \hline
Anderson & 0.557617 & 0.655576 & 1.00 \\
 \hline
MBL & 0.491745 & 0.883465 & 1.00075 \\
 \hline
TFIM chaotic & 1.00781 & 0.884371  & 0.999999 \\
 \hline
GUE & 0.999992 & 1.76251 & 2.00016 \\ [1ex] 
    \hline
\end{tabular}
\caption{The decay rate \(\gamma\) for various
  Hamiltonian models at \(\beta = 0,1,\infty\), with respect to the Ansatz \(\overline{G^{(r)}_{\beta}(t)} = \alpha
d^{-\gamma}\). The prefactor \(\alpha\) is \textit{nonuniversal}, the
details of which can be found in the Appendix.}
\label{table-parametervalues-vs-beta}
\end{table}

\prlsection{Finite-size scaling} To quantitatively understand the
numerical results, we perform finite-size scaling analysis for each
choice of \(\beta\). Let us
start with the infinite-temperature case (\(\beta=0\)). We consider an
Ansatz of the form,
\begin{align}
  \overline{G^{(r)}_{\beta}(t)} = \alpha
d^{-\gamma} + \overline{G^{(r)}_{\beta}(t)}_{\infty},
\end{align}
where \(d\) is
the Hilbert space dimension and
\(\overline{G^{(r)}_{\beta}(t)}_{\infty}\) is the asymptotic value,
i.e., as \(d \rightarrow \infty\). From several analytical and
numerical results, we know that
\(\overline{G^{(r)}_{\beta}(t)}_{\infty} = 0\). That is, the BROTOC
decays for all models, free, integrable, or chaotic. Therefore, we
reduce the Ansatz to
\begin{align}
\overline{G^{(r)}_{\beta}(t)} = \alpha
d^{-\gamma}.
\end{align}
As a result, we have,
\(\log_{2}(\overline{G^{(r)}_{\beta}(t)}) = \log_{2}(\alpha) + \left(
  -\gamma \right) L\) where \(2^{L}=d\). The numerical results naturally
manifest this Ansatz as is evident from the nearly linear figures. Therefore, performing a
linear fit to the \(\log_{2}(\overline{G^{(r)}_{\beta}(t)})\) versus
\(L\) plots yields the decay rates corresponding to various models. We
focus on the last 5 data points to obtain the fit parameters, see the
Appendix for more details.

The finite-size scaling results are summarized in
\cref{table-parametervalues-vs-beta}. The decay rates are \textit{universal} at \(\beta = 0\) with \(\gamma \approx 0.5\) for the
integrable models and \(\gamma \approx 1\) for the chaotic
models. Around \(\beta = O(1)\), this universality begins to breakdown
and at large \(\beta\), the equilibration value
\(\overline{G^{(r)}_{\beta}(t)}\) only differentiates local models
from the nonlocal GUE model.

From the analytical results about NRC-PS and GUE, we know that at
\(\beta=0\), the \(\gamma_{\mathrm{NRC\text{-}PS}} = \frac{1}{2}
\gamma_{\mathrm{GUE}}\). And, from the finite-size scaling
results, we obtain,
\(\gamma_{\mathrm{NRC\text{-}PS}}/\gamma_{\mathrm{GUE}}\approx
0.495831\). At \(\beta \rightarrow \infty\), using
\autoref{prop:zero-temperature}, for both NRC-PS and GUE, the
equilibration value is determined by the ground state
purity. Therefore, for NRC-PS, it scales as \(\frac{1}{d}\) and for
GUE it scales as \(\frac{1}{d^{2}}\) since GUE ground states are
Haar-random states, their purity is near minimum, with
\(O(\frac{1}{d})\) corrections, which is also consistent with the
finite-size scaling results. Therefore, the ratio of the rates in
this case is also \(\frac{1}{2}\). And finally, from the numerical
values listed in \cref{table-parametervalues-vs-beta}, we see that the
ratio is \(\approx \frac{1}{2}\) at \(\beta=1\) as well.

\begin{widetext}

  \begin{figure}[!th]
     \centering
 \includegraphics[width=0.8\textwidth]{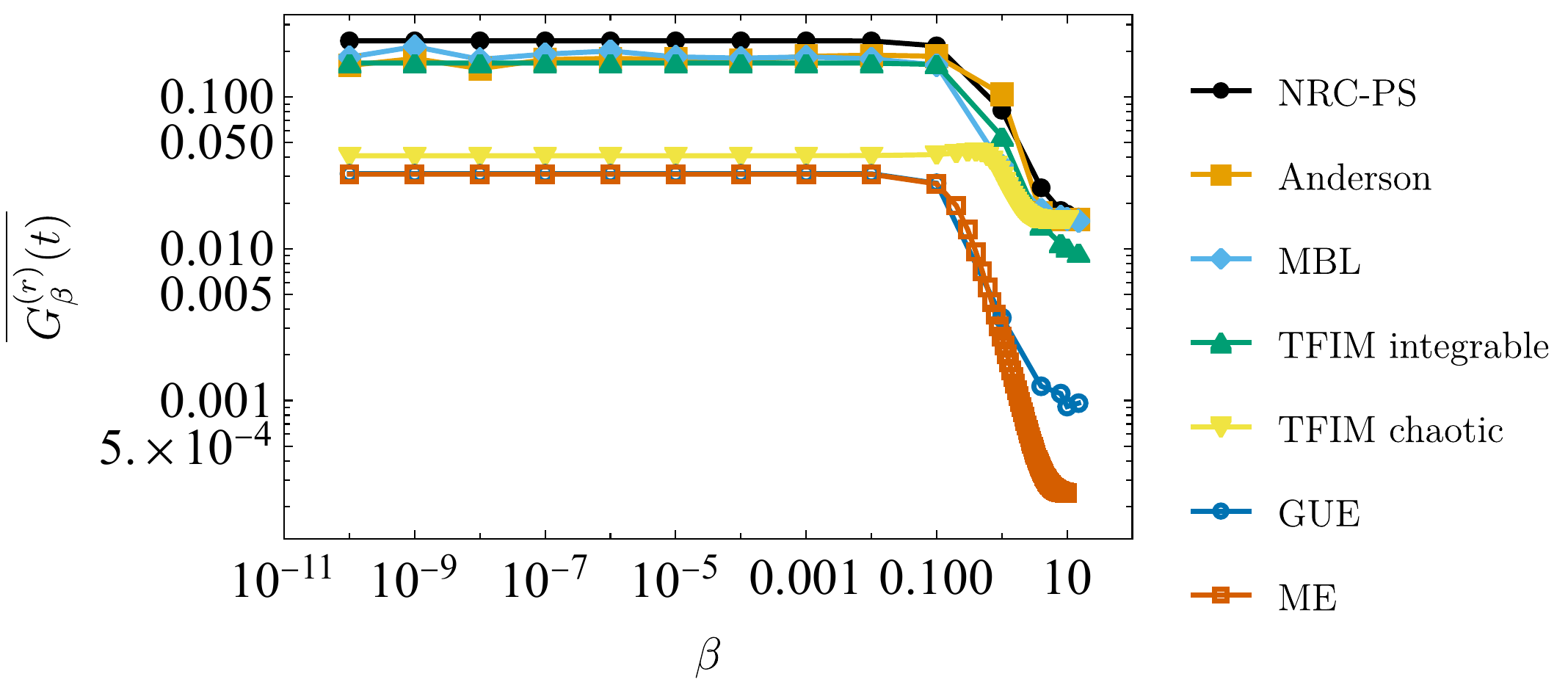}
\caption{A log-log plot of the equilibration value (long-time average) of the \(\overline{G}_{\beta}^{(r)}(t)\) for various Hamiltonian models at $L=6$ as a function of the inverse temperature \(\beta\) across a symmetric bipartition \(L/2:L/2\). We use exact time evolution for the integrable TFIM and Anderson since they do not satisfy NRC. For Anderson, MBL, and GUE, we perform exact diagonalization of the full Hamiltonian and use the analytical expression in \autoref{prop:longtimeavg}. For NRC-PS and ME (maximally entangled model), we use the analytical expressions in \cref{prop:brotoc-maxent} and \cref{G_NRC-PS}.}
\label{fig:longtimeavg-vs-beta}
\end{figure}

\begin{figure}[!th]
\centering
\includegraphics[width=0.7\textwidth]{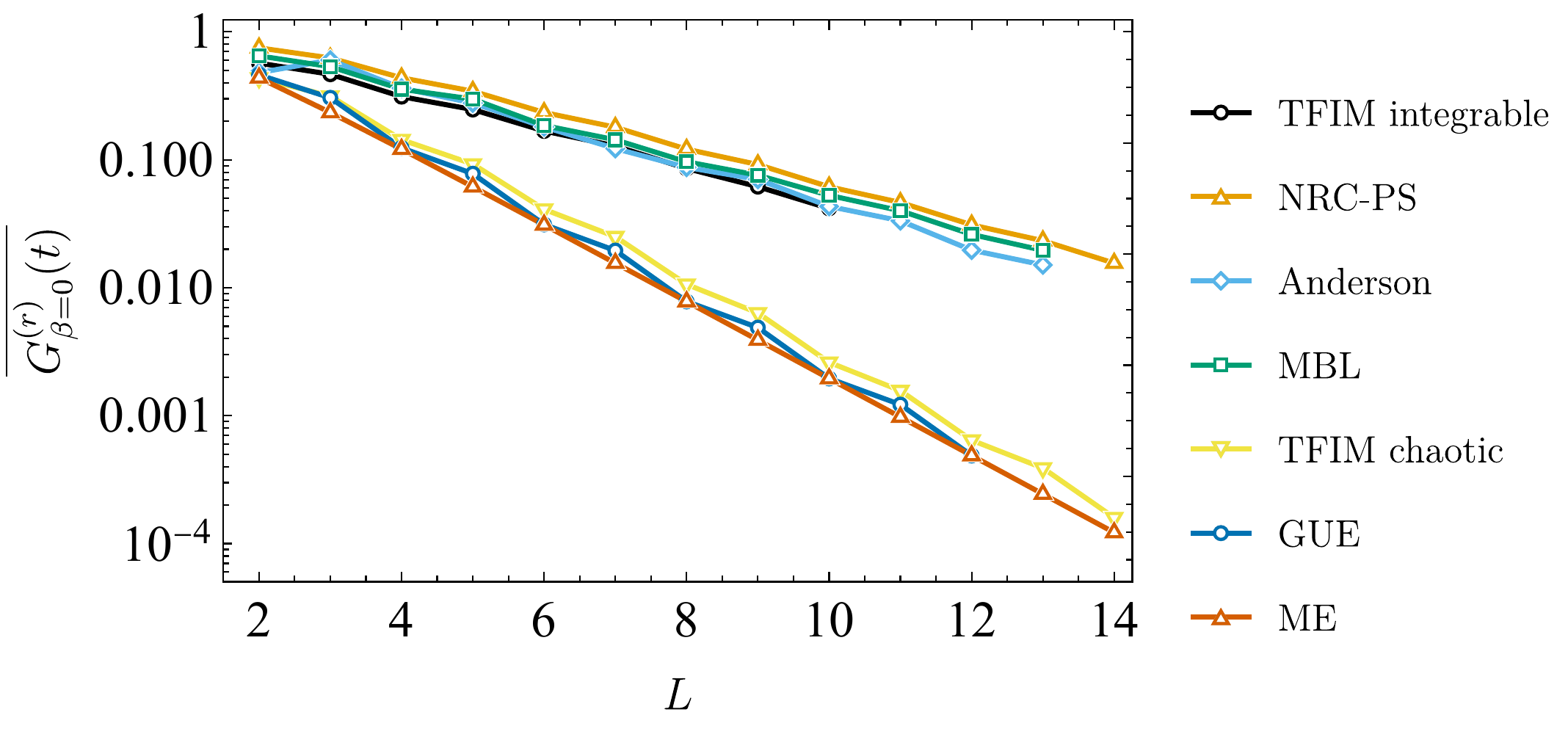}
\caption{A logplot of the equilibration value (long-time average) of the \(\overline{G}_{\beta}^{(r)}(t)\) for various Hamiltonian models as a function of the system size \(L\) at \(\beta = 0\) across a symmetric bipartition \(\floor*{L/2}:\ceil*{L/2}\). We use exact time evolution for the integrable TFIM and Anderson since they do not satisfy NRC. For Anderson, MBL, and GUE, we perform exact diagonalization of the full Hamiltonian and use the analytical expression in \autoref{prop:longtimeavg}. For NRC-PS and ME (maximally entangled model), we use the analytical expressions in \cref{prop:brotoc-maxent} and \cref{G_NRC-PS}.}
\label{fig:longtimeavg-beta-zero}
\end{figure}

\begin{figure}[!th]
\centering
\includegraphics[width=0.7\textwidth]{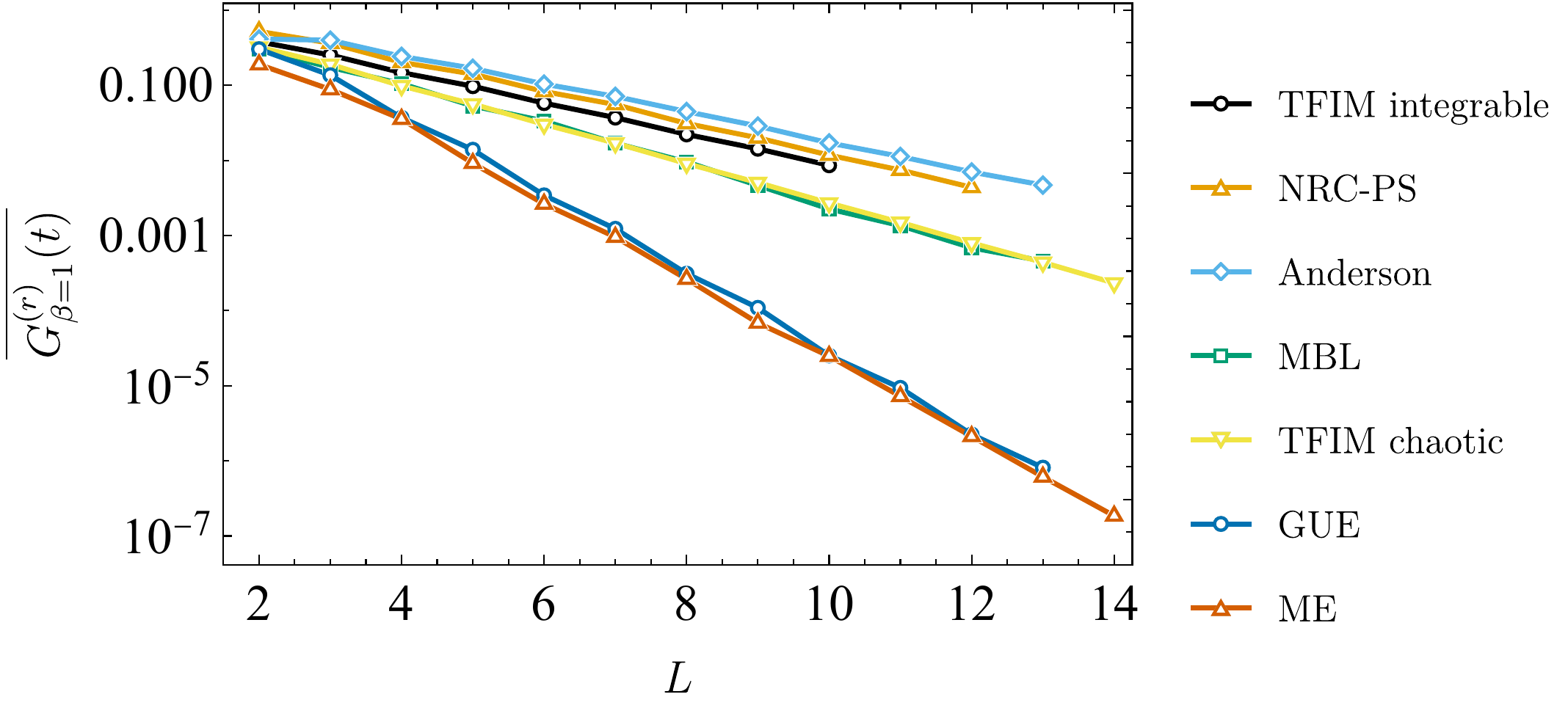}
\caption{A logplot of the equilibration value (long-time average) of the \(\overline{G}_{\beta}^{(r)}(t)\) for various Hamiltonian models as a function of the system size \(L\) at \(\beta = 1\) across a symmetric bipartition \(\floor*{L/2}:\ceil*{L/2}\). We use exact time evolution for the integrable TFIM and Anderson since they do not satisfy NRC. For Anderson, MBL, and GUE, we perform exact diagonalization of the full Hamiltonian and use the analytical expression in \autoref{prop:longtimeavg}. For NRC-PS and ME (maximally entangled model), we use the analytical expressions in \cref{prop:brotoc-maxent} and \cref{G_NRC-PS}.}
\label{fig:longtimeavg-beta-one}
\end{figure}

\begin{figure}[!th]
\centering
\includegraphics[width=0.7\textwidth]{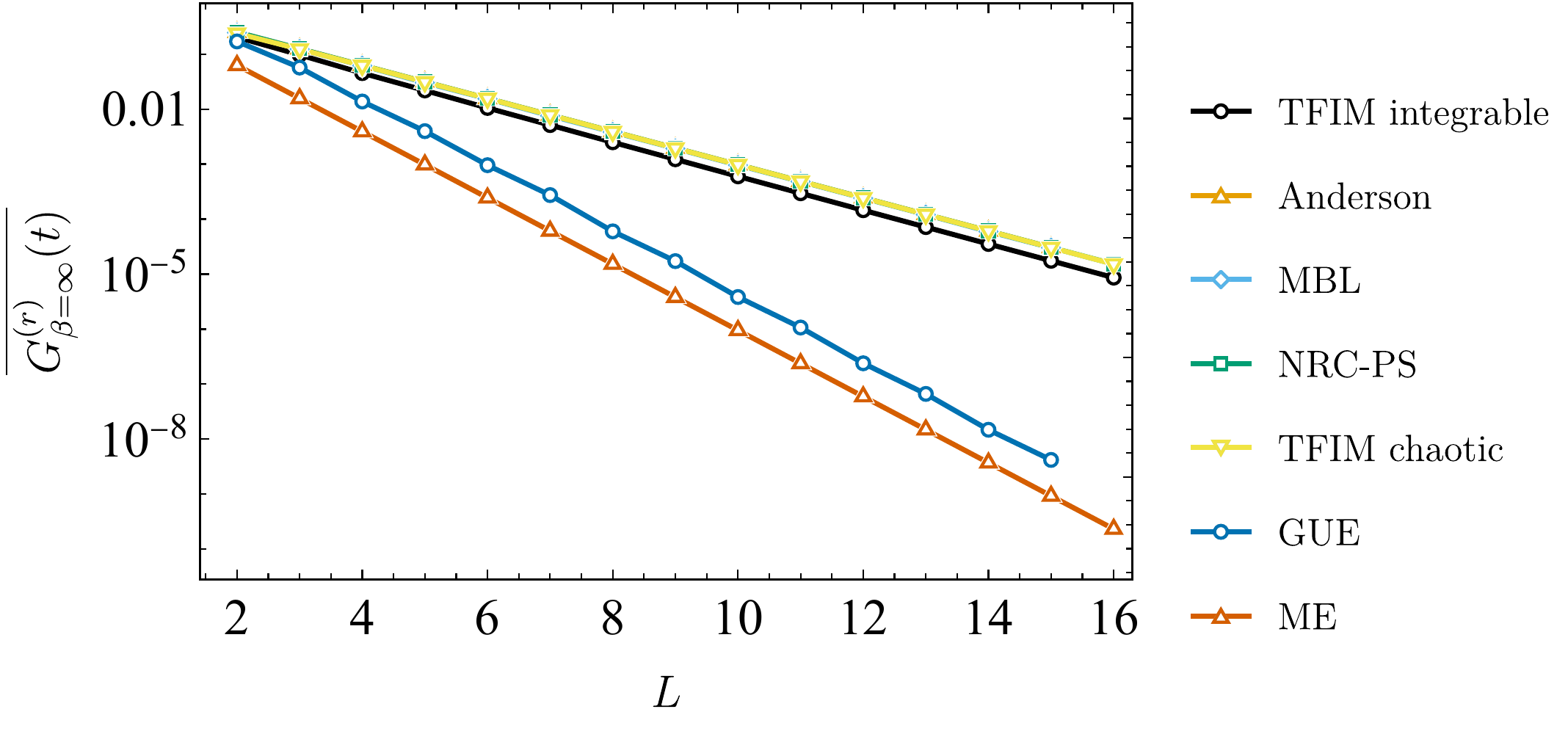}
\caption{A logplot of the equilibration value (long-time average) of the \(\overline{G}_{\beta}^{(r)}(t)\) for various Hamiltonian models as a function of the system size \(L\) at \(\beta = \infty\) across a symmetric bipartition \(\floor*{L/2}:\ceil*{L/2}\). The various data points have coalesced into two curves, first consisting of all the integrable models, whose ground state follow area-law entanglement. And second, for the GUE and ME (maximally entangled model), whose ground states follow volume-law entanglement. Using \autoref{prop:zero-temperature}, we simply compute the ground state projector for various models to compute this numerically.}
\label{fig:longtimeavg-beta-inf}
\end{figure}
\end{widetext}

\section{Conclusions}
\label{sec:conclusions}
In this work we introduce the bipartite regularized OTOC that
allows us to obtain a wealth of analytical and numerical results to
aid our understanding of regularized OTOCs for local quantum
systems. The infinite-temperature OTOC has several operational
interpretations in terms of operator entanglement, entropy production, and
others. Proposition 1 established the
connected component of the BROTOC as probing the operator purity of a quantum
operation. We then showed that the quantum operation
\(\mathcal{U}_{\beta,t}\) is intimately related to the two-point spectral form
factor and globally averaged regularized OTOCs are connected to
the four-point spectral form factor, respectively. Moreover, the
connected BROTOC probes the purity of the associated thermofield
double state.

Moving away from the infinite-temperature assumption, we investigate
the zero-temperature case, where, in
Proposition 3, we showed that, in this limit, both the
disconnected and connected components of the BROTOC probe the
groundstate entanglement for nondegenerate Hamiltonians. This allows
us to think of them as probes to quantum phase transitions in the system.

In Propositions 4 and 5, we study the equilibration value of the BROTOC
and how it connects to \textit{eigenstate entanglement}. In fact, we
show that if there is sufficient entanglement in all eigenstates
across the spectrum, then the equilibration value must be nearly maximal. We also obtain analytical closed-form
expressions for the equilibration value of nearly maximally-entangled
Hamiltonians and contrast the \(\beta\)-dependence in the
unregularized versus the regularized case.

Finally, we perform numerical simulations on various integrable and
chaotic Hamiltonian models to study the equilibration value of the
connected component of the BROTOC. Using a mix of finite-size scaling
and analytical estimates, we contrast the decay rates of the BROTOC
for various models. While at \(\beta = 0\), the decay rate is
universal and distinguishes integrable, chaotic, and random matrix
evolutions; as we reach \(\beta = O(1)\), this universality begins to
break down. And, in fact, at \(\beta \rightarrow \infty\), the
equilibration value only distinguishes local models from the GUE, and
is therefore no longer a reliable signature  of chaotic-vs-integrable
dynamics. An interesting future work would be to contrast various
choices of regularizations and their ability to distinguish chaotic
and integrable dynamics.\\

\section{Acknowledgments}
N.A. would like to thank G. Styliaris for many insightful
discussions. The authors acknowledge the Center for Advanced Research Computing
(CARC) at the University of Southern California for providing
computing resources that have contributed to the research results
reported within this publication. URL: \url{https://carc.usc.edu}. The
authors acknowledge partial support from the NSF award PHY-1819189. This research was (partially) sponsored by the Army Research Office and was accomplished under Grant Number W911NF-20-1-0075. The views and conclusions contained in this document are those of the authors and should not be interpreted as representing the official policies, either expressed or implied, of the Army Research Office or the U.S. Government. The U.S. Government is authorized to reproduce and distribute reprints for Government purposes notwithstanding any copyright notation herein.

\bibliographystyle{apsrev4-1}
\bibliography{my_library,pra_referee}

\onecolumngrid
\widetext
\newpage

\appendix

\renewcommand{\thepage}{A\arabic{page}}
\setcounter{page}{1}
\renewcommand{\thesection}{A\arabic{section}}
\setcounter{section}{0}
\renewcommand{\thetable}{A\arabic{table}}
\setcounter{table}{0}
\renewcommand{\theequation}{A\arabic{equation}}
\setcounter{equation}{0}

\section*{Appendices}
\label{sec:appendix}

\subsection*{Proof of \autoref{prop:regularized-otoc-main}}
\regularizedotocmain*

Consider a bipartite Hilbert space, \(\mathcal{H}_{AB} =
\mathcal{H}_{A} \otimes \mathcal{H}_{B}\). Let \(V \in
\mathcal{U}(\mathcal{H}_{A}), W \in \mathcal{U}(\mathcal{H}_{B})\) and
\(\mathbb{E}_{U \in \mathcal{U}(\mathcal{H})} \left[ f(U) \right]\)
denote the Haar average of \(f(U)\) over
\(\mathcal{U}(\mathcal{H})\). Then, using the lemma \cite{watrousTheoryQuantumInformation2018},
\begin{align}
\mathbb{E}_{U \in \mathcal{U}(\mathcal{H})} \left[ U \otimes U^{\dagger} \right]  = \frac{\mathbb{S}}{d},
\end{align}
where \(\mathbb{S}\) is the swap operator on \(\mathcal{H} \otimes
\mathcal{H}'\) with \(\mathcal{H}'\) representing a replica of the
original Hilbert space \(\mathcal{H}\). Given an orthonormal basis of
\(\mathcal{H}\), \(\mathbb{B} = \{ | j \rangle \}_{j=1}^{d}\) (and a
replica of the basis for \(\mathcal{H}'\)), the swap operator can be
represented as \(\mathbb{S} = \sum\limits_{j,k=1}^{d} | j \rangle
\langle  k | \otimes | k \rangle \langle  j |\). Now, if we consider
Haar averages over a subsystem instead, then the lemma above is modified as,
\begin{align}
\mathbb{E}_{V \in \mathcal{U}(\mathcal{H}_{A})} V^{\dagger} \otimes V
  = \frac{\mathbb{S}_{AA'}}{d_{A}} \text{ and } \mathbb{E}_{W \in \mathcal{U}(\mathcal{H}_{B})} W^{\dagger} \otimes W = \frac{\mathbb{S}_{BB'}}{d_{B}}.
\end{align}

First, we compute the disconnected BROTOC,
\begin{align}
G^{(d)}_{\beta} &:= \mathbb{E}_{V \in \mathcal{U}(\mathcal{H}_{A}), W \in \mathcal{U}(\mathcal{H}_{B})} F^{(d)}_{\beta} \\
                &= \mathbb{E}_{V \in \mathcal{U}(\mathcal{H}_{A}), W \in \mathcal{U}(\mathcal{H}_{B})} \operatorname{Tr}\left[ \left( \sqrt{\rho_{\beta}} \otimes \sqrt{\rho_{\beta}} \right) \left( W^{\dagger} \otimes W \right) \mathbb{S} \right] \operatorname{Tr}\left[ \left( \sqrt{\rho_{\beta}} \otimes \sqrt{\rho_{\beta}} \right) \left( V^{\dagger} \otimes V \right) \mathbb{S} \right] \\
                &=\frac{1}{d} \operatorname{Tr}\left[ \mathbb{S} \left( \sqrt{\rho_{\beta}} \right)^{\otimes 2} \mathbb{S}_{BB'} \right]  \operatorname{Tr}\left[ \mathbb{S} \left( \sqrt{\rho_{\beta}} \right)^{\otimes 2} \mathbb{S}_{AA'} \right] = \frac{1}{d} \operatorname{Tr}\left[ \left( \sqrt{\rho_{\beta}} \right)^{\otimes 2} \mathbb{S}_{AA'} \right] \operatorname{Tr}\left[ \left( \sqrt{\rho_{\beta}} \right)^{\otimes 2} \mathbb{S}_{BB'} \right],
\end{align}
where in the second line, we have used the lemma, \(\operatorname{Tr}\left[ \left( A \otimes B \right) \mathbb{S} \right] = \operatorname{Tr}\left[ AB \right]\) and in the third line we have used \(\mathbb{S} = \mathbb{S}_{AA'} \mathbb{S}_{BB'}\), \(\mathbb{S}^{2} = \mathbb{I}\), \(\mathbb{S}_{AA'}^{2} = \mathbb{I}_{AA'}\), and \(\mathbb{S}_{BB'}^{2} = \mathbb{I}_{BB'}\).

We can formally perform partial traces above, for example,
\begin{align}
\operatorname{Tr}\left[ \left( \sqrt{\rho_{\beta}} \otimes \sqrt{\rho_{\beta}} \right) \mathbb{S}_{AA'} \right] = \operatorname{Tr}_{AA'}\left[ \operatorname{Tr}_{BB'}\left[  \left( \sqrt{\rho_{\beta}} \otimes \sqrt{\rho_{\beta}} \right) \right] S_{AA'} \right] = \operatorname{Tr}_{AA'}\left[ \sigma_{A} \otimes \sigma_{A'}\mathbb{S}_{AA'} \right] = \operatorname{Tr}\left[ \sigma_{A}^{2} \right],
\end{align}
where \(\sigma_{A} \equiv \operatorname{Tr}_{B}\left[
  \sqrt{\rho_{\beta}} \right]\) above for brevity. Let
\(\mathcal{P}_{\chi}(\rho):= \left\Vert \rho_{\chi}
\right\Vert_{2}^{2}\) be the squared \(2\)-norm of the operator
\(\rho_{\chi}\) with \(\rho_{\chi}:=
\operatorname{Tr}_{\overline{\chi}}\left[ \rho \right]\) and \(\chi =
\{ A,B \}\) with \(\overline{chi}\) the complement of \(\chi\). Then,
\begin{align}
G_{d}(\beta) = \frac{1}{d} \mathcal{P}_{A}(\sqrt{\rho_{\beta}}) \mathcal{P}_{B}(\sqrt{\rho_{\beta}}).
\end{align}

Similarly, for the connected BROTOC we have,
\begin{align}
G^{(r)}_{\beta}(t) &:= \mathbb{E}_{V \in \mathcal{U}(\mathcal{H}_{A}), W \in \mathcal{U}(\mathcal{H}_{B})} F^{(r)}_{\beta}(t) \\
&= \mathbb{E}_{V \in \mathcal{U}(\mathcal{H}_{A}), W \in \mathcal{U}(\mathcal{H}_{B})} \operatorname{Tr}\left[ \mathbb{S} \left( W^{\dagger}_{t} \otimes W_{t} \right) \left( y \otimes y \right) \left( V^{\dagger} \otimes V \right) \left( y \otimes y \right) \right]\\
&= \frac{1}{d} \operatorname{Tr}\left[ \mathbb{S}_{AA'} y^{\otimes 2} U^{\dagger \otimes 2}_{t} S_{AA'} U^{\otimes 2}_{t}  y^{\otimes 2} \right]\\
&= \frac{1}{d \mathcal{Z}(\beta)} \operatorname{Tr}\left[ \mathbb{S}_{AA'} \mathcal{U}_{\beta,t}^{\otimes 2} \left( \mathbb{S}_{AA'} \right) \right],
\end{align}
where  \(\mathcal{U}_{\beta,t}:= \mathcal{V}_{\beta} \circ
\mathcal{U}_{t}\) with \(\mathcal{V}_{\beta}(X):= \exp \left[ -\beta
  H/4 \right] X \exp \left[ -\beta H/4 \right] \) the imaginary time-evolution,
\(\mathcal{U}_{t}(X):= U^{\dagger}_{t} X U_{t}\) the real
time-evolution, and \(U_{t} = \exp
\left[ -iHt \right] \) the usual time-evolution operator. This completes the proof.

\subsection*{Proof of \autoref{prop:longtimeavg}}
\longtimeaverage*

Recall that NRC implies nondegeneracy of the spectrum, therefore,
using the spectral decomposition of a Hamiltonian, \(H =
\sum\limits_{j=1}^{d} E_{j} \Pi_{j}\) with \(\Pi_{j} = | \phi_{j}
\rangle \langle  \phi_{j} | \), we have,
\begin{align}
\mathcal{U}_{\beta,t}^{\otimes 2}(A) = \sum\limits_{j,k,l,m}^{d} \exp \left[ -\beta/4 \left( E_{j} + E_{k} + E_{l}  +E_{m} \right) - it \left( E_{j} + E_{k}  - E_{l} - E_{m}\right) \right] \left( \Pi_{j} \otimes \Pi_{k} \right) A \left( \Pi_{l} \otimes \Pi_{m} \right).
\end{align}

Using the NRC assumption, we have the following, \(\overline{\exp \left[ - it \left( E_{j} + E_{k}  - E_{l} - E_{m}\right) \right] }^{t} = \delta_{jl} \delta_{km} + \delta_{jm} \delta_{kl} - \delta_{jk} \delta_{kl} \delta_{lm}\). Therefore,
\begin{align}
\overline{G^{(r)}_{\beta}(t)} = \frac{1}{d \mathcal{Z}(\beta)} & \left(  \sum\limits_{j,k}^{d} \exp \left[ -\beta/2 \left( E_{j} + E_{k} \right)\right] \operatorname{Tr}\left[ \left( \Pi_{j} \otimes \Pi_{k} \right) \mathbb{S}_{AA'} \left( \Pi_{j} \otimes \Pi_{k} \right) \mathbb{S}_{AA'}\right] \right. \\ &+ \sum\limits_{j,k}^{d} \exp \left[ -\beta/2 \left( E_{j} + E_{k} \right)\right] \operatorname{Tr}\left[ \left( \Pi_{j} \otimes \Pi_{k} \right) \mathbb{S}_{BB'} \left( \Pi_{j} \otimes \Pi_{k} \right) \mathbb{S}_{BB'}\right] \\ &- \left. \sum\limits_{j}^{d} \exp \left[ -\beta E_{j} \right] \operatorname{Tr}\left[ \Pi_{j}^{\otimes 2} \mathbb{S}_{AA'} \Pi_{j}^{\otimes 2} \mathbb{S}_{AA'}\right] \right).
\end{align}
Now, for pure states \(\Pi_{j}, \Pi_{k}\) one can show that
\(\operatorname{Tr}\left[ \left( \Pi_{j} \otimes \Pi_{k} \right)
  \mathbb{S}_{AA'} \left( \Pi_{j} \otimes \Pi_{k} \right)
  \mathbb{S}_{AA'}\right] = \left| \operatorname{Tr}\left[ \left(
      \Pi_{j} \otimes \Pi_{k} \right) \mathbb{S}_{AA'} \right]
\right|^{2}\).

We now define the \(R\)-matrix introduced in
Ref. \cite{styliaris_information_2021} while computing the
infinite-temperature variant of this Proposition. Let \(\rho_{j}^{A(B)}:=
\operatorname{Tr}_{A(B)}\left[ \Pi_{j} \right]\) be the reduced state,
then, formally performing the partial trace we have,
\begin{align}
\operatorname{Tr}_{AA'BB'}\left[ \Pi_{k} \otimes \Pi_{l} S_{AA'} \right] = \operatorname{Tr}_{AA'}\left[ \operatorname{Tr}_{BB'}\left[ \Pi_{k} \otimes \Pi_{l} \right] S_{AA'} \right] = \operatorname{Tr}_{AA'}\left[ \rho_{k}^{A} \otimes \rho_{l}^{A'} S_{AA'} \right] = \operatorname{Tr}\left[ \rho_{k}^{A} \rho_{l}^{A} \right] =: R_{kl}^{A}.
\end{align}
Therefore, the time-average can be simplified as,
\begin{align}
\overline{G_{\beta}^{(r)}(t)} = \frac{1}{d\mathcal{Z}(\beta)} \left( \sum\limits_{j,k}^{d} \exp \left[ -\beta/2 (E_{j} + E_{k}) \right] \left( \left| R_{jk}^{A} \right|^{2} + \left| R_{jk}^{B} \right|^{2} - \delta_{jk} \left| R_{jk}^{A} \right|^{2} \right) \right).
\end{align}

Now we introduce a more compact notation for the equilibration
value. Let \(\tilde{R}^{\chi}_{jk}:= \exp \left[ -\beta \left( E_{j} +
  E_{k}\right)/4 \right] \left\langle \rho_{j}^{\chi}, \rho_{k}^{\chi}
\right\rangle\), then,
\begin{align}
\overline{G_{\beta}^{(r)}(t)} = \frac{1}{d \mathcal{Z}(\beta)} \sum\limits_{\chi \in \{ A,B \}}^{} \left( \left\Vert \widetilde{R}^{(\chi)} \right\Vert_{2}^{2} - \frac{1}{2} \left\Vert \widetilde{R}_{D}^{(\chi)} \right\Vert_{2}^{2} \right),
\end{align}
with \([ \widetilde{R}_{D}^{(\chi)} ]_{jk} = [
\widetilde{R}_{D}^{(\chi)} ]_{jk}  \delta_{jk}\); where we have used the fact that \(R_{kk}^{A} = R_{kk}^{B}\), that is,
the the reduced states \(\rho_{j}^A\) and \(\rho_{j}^{B}\) are
isospectral (up to irrelevant zeros).

\subsection*{Proof of \autoref{prop:eigenstate-entanglement}}
\eigenstateentanglement*

We prove this for the general case \(d_{A} \neq d_{B}\), and the original Proposition can be recovered by setting \(d_{A} = d_{B}\) at the end. Let us assume without loss of generality that \(d_{A} \leq
d_{B}\). The spectral decomposition of the Hamiltonian is \(H =
\sum\limits_{j=1}^{d} E_{j} | \phi_{j} \rangle \langle  \phi_{j} |
\) and its reduced states are labelled as \(\rho_{j}^{A}:=
\operatorname{Tr}_{B}\left[ | \phi_{j} \rangle \langle  \phi_{j} |
\right]\).

Notice that since \(\mathcal{P}(| \psi_{AB} \rangle) - \mathcal{P}_{\min}
= \operatorname{Tr}\left[ \rho_{A}^{2} \right] -
\frac{1}{d_{A}}\), the assumption that the purities are nearly minimum can be equivalently expressed as: \(\mathcal{P}(\rho_{k}^{A}) - \mathcal{P}_{\min} \leq \epsilon \iff
\left\Vert \Delta_{k}^{A} \right\Vert_{2}^{2} \leq \epsilon\) with \(\Delta_{k}^{A}:=\rho_{k}^{A}-I /
d_{A} ~~\forall k\).

Define \(\Delta_{k}^{B}\) analogously, i.e., \(\Delta_{k}^{B}:=\rho_{k}^{B}-I /
d_{B}\). Then, using the fact that the reduced states of a pure state are
isospectral (therefore, \(\mathcal{P}(\rho_{k}^{A}) = \mathcal{P}(\rho_{k}^{B}) ~~\forall k\)) and \(d_{A} \leq d_{B}\), we have, \(\left\Vert \Delta_{k}^{B}
\right\Vert_{2}^{2} = \left\Vert \rho_{k}^{B} \right\Vert_{2}^{2} -
\frac{1}{d_{B}} =  \left\Vert \rho_{k}^{A} \right\Vert_{2}^{2} -
\frac{1}{d_{B}} \leq \left\Vert \rho_{k}^{A} \right\Vert_{2}^{2} -
\frac{1}{d_{A}} \leq \epsilon\) \(~~\forall k\).

Now, recall that, \(\overline{G_{\beta}^{(r)}(t)} |_{\mathrm{ME}} =
\frac{1}{d^{2}}\left(\frac{2 \mathcal{Z}(\beta /
    2)^{2}}{\mathcal{Z}(\beta)}-1\right)\) and \(\overline{G^{(r)}_{\beta}(t)} = \frac{1}{d \mathcal{Z}(\beta)}
\left( \left\Vert \tilde{R}^{A} \right\Vert_{2}^{2} + \left\Vert \tilde{R}^{B} \right\Vert_{2}^{2} - \left\Vert \tilde{R}^{A}_{D} \right\Vert_{2}^{2} \right)\), where we have used the fact that \(\left\Vert
  \tilde{R}_{D}^{(A)} \right\Vert_{2}^{2} = \left\Vert \tilde{R}_{D}^{(B)}
\right\Vert_{2}^{2}\). Then,
\begin{align}
& \left| \overline{G_{\beta}^{(r)}(t)} |_{\mathrm{ME}} - \overline{G_{\beta}^{(r)}(t)} |_{\mathrm{NRC}} \right| =  \left| \frac{1}{d^{2}} \left( \frac{2 \mathcal{Z}^{2}(\beta/2)}{\mathcal{Z}(\beta)} - 1 \right) - \frac{1}{d \mathcal{Z}(\beta)} \left( \left\Vert \widetilde{R}^{(A)} \right\Vert_{2}^{2} + \left\Vert \widetilde{R}^{(B)} \right\Vert_{2}^{2} - \left\Vert \widetilde{R}_{D}^{(A)} \right\Vert_{2}^{2} \right) \right|\\
  & \leq \underbrace{\left| \frac{1}{d^{2} \mathcal{Z}(\beta)} \mathcal{Z}^{2}(\beta/2)  - \frac{1}{d \mathcal{Z}(\beta)} \left\Vert \widetilde{R}^{(A)} \right\Vert_{2}^{2} \right|}_{(i)}
    + \underbrace{ \left| \frac{1}{d^{2} \mathcal{Z}(\beta)} \mathcal{Z}^{2}(\beta/2)  - \frac{1}{d \mathcal{Z}(\beta)} \left\Vert \widetilde{R}^{(B)} \right\Vert_{2}^{2} \right|}_{(ii)}
    + \underbrace{\left| \frac{1}{d \mathcal{Z}(\beta)} \left\Vert \widetilde{R}^{(A)}_{D} \right\Vert_{2}^{2} - \frac{1}{d^{2}} \right|}_{(iii)},
\end{align}
where we have split the terms in \(\overline{G_{\beta}^{(r)}(t)} |_{\mathrm{ME}}\) and then used the triangle inequality. We now want to bound each of the terms \((i),(ii), \text{ and } (iii)\).

Since \(\left\Vert \widetilde{R}^{(\chi)} \right\Vert_{2}^{2} =
\sum\limits_{j,k=1}^{d} \exp \left[ -\beta \left( E_{j} + E_{k}
  \right)/2 \right] \left( \left\langle \rho_{j}^{(\chi)}, \rho_{k}^{(\chi)}
  \right\rangle \right)^{2}\) (note that we do not need an absolute value here since the exponential term is nonnegative and \(A,B \geq 0 \implies \left\langle A,B \right\rangle \geq 0\) so the inner products between the reduced states is nonnegative as well); we need to bound \(\left( \left\langle \rho_{j}^{(\chi)}, \rho_{k}^{(\chi)}
  \right\rangle \right)^{2}\). Note that,
\begin{align}
\left(\left\langle\rho_{k}^{(\chi)}, \rho_{l}^{(\chi)}\right\rangle\right)^{2}=\left(\left\langle I / d_{\chi}+\Delta_{k}^{(\chi)}, I / d_{\chi}+\Delta_{l}^{(\chi)}\right\rangle\right)^{2}=\left(\frac{1}{d_{\chi}}+\left\langle\Delta_{k}^{(\chi)}, \Delta_{l}^{(\chi)}\right\rangle\right)^{2}=\frac{1}{d_{\chi}^{2}}+\frac{2}{d_{\chi}}\left\langle\Delta_{k}^{(\chi)}, \Delta_{l}^{(\chi)}\right\rangle+\left\langle\Delta_{k}^{(\chi)}, \Delta_{l}^{(\chi)}\right\rangle^{2}.
\end{align}
And, using Cauchy-Schwarz inequality along with the fact
that \(\left\langle \Delta_{k}^{\chi},
  \Delta_{k}^{\chi} \right\rangle = \left\Vert \rho_{k}^{(\chi)}
\right\Vert_{2}^{2} - \frac{1}{d_{\chi}} \leq \epsilon ~~\forall
k\) (as shown above), we have, \(\left\langle \Delta_{k}^{\chi},
  \Delta_{l}^{\chi} \right\rangle \leq \sqrt{\left\langle \Delta_{k}^{\chi},
  \Delta_{k}^{\chi} \right\rangle \left\langle \Delta_{l}^{\chi},
  \Delta_{l}^{\chi} \right\rangle } \leq \epsilon ~~\forall k,l\). Therefore, \(\left(\left\langle\rho_{k}^{(\chi)},
    \rho_{l}^{(\chi)}\right\rangle\right)^{2} \leq
\frac{1}{d_{\chi}^{2}} + \frac{2 \epsilon}{d_{\chi}} +
\epsilon^{2} \equiv f(d_{\chi},\epsilon)\) \(~~\forall k,l\). 

Plugging this back in \(\left\Vert
  \tilde{R}^{(\chi)} \right\Vert_{2}^{2}\), we have, \(\left\Vert \widetilde{R}^{(\chi)} \right\Vert_{2}^{2} =
\sum\limits_{j,k=1}^{d} \exp \left[ -\beta \left( E_{j} + E_{k}
\right)/2 \right] f(d_{\chi},\epsilon) = \mathcal{Z}(\beta/2)^{2} f(d_{\chi},\epsilon)\). Now, for the diagonal part, \(\left\Vert \tilde{R}_{D}^{A} \right\Vert_{2}^{2} = \sum\limits_{j=1}^{d} \exp \left[ -\beta E_{j} \right] \left( \left\langle \rho_{j}^{A}, \rho_{j}^{A} \right\rangle \right)^{2}  \leq \sum\limits_{j=1}^{d} \exp \left[ -\beta E_{j} \right] f(d_{A},\epsilon) = \mathcal{Z}(\beta) f(d_{A},\epsilon)\).

Therefore, term \((i)\) above becomes \(\frac{\mathcal{Z}^{2}(\beta/2)}{d \mathcal{Z}(\beta)} \left| \frac{1}{d} - f(d_{A},\epsilon) \right|\) and term \((ii)\) becomes, \(\frac{\mathcal{Z}^{2}(\beta/2)}{d \mathcal{Z}(\beta)} \left| \frac{1}{d} - f(d_{B},\epsilon) \right|\). And, term \((iii)\) becomes, \(\frac{1}{d} \left| \frac{1}{d} - f(d_{A},\epsilon) \right|\). Now, notice that 
\begin{align}
\mathcal{Z}(\beta/2)^{2} &= \sum\limits_{j,k=1}^{d} \exp \left[ -\beta E_{j}/2 \right] \exp \left[ -\beta E_{k}/2 \right] = \sum\limits_{j=k}^{d} \exp \left[ -\beta E_{j}/2 \right] \exp \left[ -\beta E_{k}/2 \right] \\
&+ \sum\limits_{j \neq k}^{d} \exp \left[ -\beta E_{j}/2 \right] \exp \left[ -\beta E_{k}/2 \right] \geq \sum\limits_{j=1}^{d} \exp \left[ -\beta E_{j} \right] = \mathcal{Z}(\beta),
\end{align}
where we have dropped the \(j \neq k\) terms in  the summation and used their nonnegativity. Therefore, \(\mathcal{Z}(\beta/2)^{2}/\mathcal{Z}(\beta) \geq 1\) and so, term (iii) can be upper bounded as \(\frac{1}{d} \left| \frac{1}{d} - f(d_{A},\epsilon) \right| \leq  \frac{\mathcal{Z}(\beta/2)^{2}}{d \mathcal{Z}(\beta)} \left| \frac{1}{d} - f(d_{A},\epsilon) \right|\).

Putting everything together, we have
\begin{align}
\left| \overline{G_{\beta}^{(r)}(t)} |_{\mathrm{ME}} - \overline{G_{\beta}^{(r)}(t)} |_{\mathrm{NRC}} \right| \leq \frac{\mathcal{Z}(\beta/2)^{2}}{d \mathcal{Z}(\beta)} \left( 2 \left| \frac{1}{d} - f(d_{A},\epsilon) \right| + \left| \frac{1}{d} - f(d_{B},\epsilon) \right| \right).
\end{align}

Now, if we set \(d_{A} = d_{B} = \sqrt{d}\), we have, \(\left| \frac{1}{d} - f(d_{\chi,\epsilon}) \right| = \left| \frac{1}{d} - \left( \frac{1}{d} + \frac{2 \epsilon}{\sqrt{d}} + \epsilon^{2} \right) \right| = \left( \frac{2 \epsilon}{\sqrt{d}} + \epsilon^{2} \right)\). Therefore,
\begin{align}
\left| \overline{G_{\beta}^{(r)}(t)} |_{\mathrm{ME}} - \overline{G_{\beta}^{(r)}(t)} |_{\mathrm{NRC}} \right| \leq \frac{\mathcal{Z}(\beta/2)^{2}}{d \mathcal{Z}(\beta)}  \left( \frac{6 \epsilon}{\sqrt{d}} + 3\epsilon^{2} \right).
\end{align}

\subsection*{Proof of \autoref{prop:sff-thermal-otoc}}
\sffthermalotoc*

We have,
\begin{align}
  F_{\beta}^{(A_{1},B_{1},A_{2},B_{2})}(t) = \frac{1}{\mathcal{Z}(\beta)}\operatorname{Tr}\left[x \mathcal{U}_{t}(A_{1}) x B_{1} x \mathcal{U}_{t} (A_{2}) x \left(  A^{\dagger}_{2} B^{\dagger}_{1} A^{\dagger}_{1} \right)  \right],
\end{align}
where we have used \(B_{2} = A^{\dagger}_{2}B^{\dagger}_{1}
A^{\dagger}_{1}\). Now, using the cyclicity of trace and the lemma
\(\mathbb{E}_{A \in \mathcal{U}(\mathcal{H})} A X A^{\dagger} =
\frac{\operatorname{Tr}\left[ X \right]}{d}\), we have,
\begin{align}
  \mathbb{E}_{A_{1} \in \mathcal{U}(\mathcal{H})}  F_{\beta}^{(A_{1},B_{1},A_{2},B_{2})}(t) &= \frac{1}{d\mathcal{Z}(\beta)} \operatorname{Tr}\left[ x U^{\dagger}_{t} \right] \operatorname{Tr}\left[ U_{t} x B_{1} x \mathcal{U}_{t} (A_{2}) x \left(  A^{\dagger}_{2} B^{\dagger}_{1} \right)  \right]\\
  &= \frac{1}{d\mathcal{Z}(\beta)} \operatorname{Tr}\left[ x U^{\dagger}_{t} \right] \operatorname{Tr}\left[ B^{\dagger}_{1} U_{t} x B_{1} x \mathcal{U}_{t} (A_{2}) x \left(  A^{\dagger}_{2}  \right)  \right],
\end{align}
where in the second line we have used the cyclicity of trace to move
\(B^{\dagger}_{1}\) to the front. Now, performing
\(\mathbb{E}_{B_{1}}\) using the lemma above, we have, 
\begin{align}
  \mathbb{E}_{B_{1} \in \mathcal{U}(\mathcal{H})}  \operatorname{Tr}\left[ B^{\dagger}_{1} U_{t} x B_{1} x \mathcal{U}_{t} (A_{2}) x \left(  A^{\dagger}_{2}  \right)  \right] = \frac{1}{d} \operatorname{Tr}\left[ U_{t}x \right] \operatorname{Tr}\left[  x U^{\dagger}_{t} A_{2} U_{t} x A^{\dagger}_{2} \right].
\end{align}
Similarly, performing the average over \(A_{2}\) we obtain the final
terms that are proportional to \(\operatorname{Tr}\left[ U_{t}x
\right] \operatorname{Tr}\left[ U^{\dagger}_{t}x \right]\). Thus, we
have the desired result.

\subsection*{Unregularized OTOCs for maximally entangled models}
Consider an operator Schmidt decomposition of \(U_{t} \equiv U =
\sum\limits_{j}^{} \sqrt{\lambda_{j}} U_{j} \otimes W_{j}\) with
\(U_{j} \in \mathcal{L}(\mathcal{H}_{A})\) and \(W_{j} \in
\mathcal{L}(\mathcal{H}_{B})\) such that \(\left\langle U_{j},U_{k}
\right\rangle = d_{A} \delta_{jk}\) and \(\left\langle W_{j}, W_{k}
\right\rangle = d_{B} \delta_{jk}\). Forr a unitary operator,
\(\left\Vert U \right\Vert_{2}^{2} = d\), therefore, we have,
\begin{align}
\left\Vert U \right\Vert_{2}^{2} &= \operatorname{Tr}\left[ \sum\limits_{j,k}^{} \sqrt{\lambda_{j} \lambda_{k}} U_{j} U^{\dagger}_{k} \otimes W_{j} W^{\dagger}_{k} \right] = \sum\limits_{j,l}^{} \sqrt{\lambda_{j} \lambda_{k}} \left\langle U_{j}, U_{k} \right\rangle \left\langle W_{j}, W_{k} \right\rangle = d_{A} d_{B}  \sum\limits_{j}^{} \lambda_{j} \implies \sum\limits_{j}^{} \lambda_{j} = 1.
\end{align}

Now, consider the bipartite unregularized OTOC, \(F_{\beta}(U) = \frac{1}{d} \operatorname{Tr}\left[ \left(
    \rho_{\beta} \otimes \mathbb{I} \right) U^{\otimes 2}
  \mathbb{S}_{AA'} U^{\dagger \otimes 2} \mathbb{S}_{AA'}
\right]\). Plugging in the operator Schmidt decomposition of \(U\), we have,
\begin{align*}
&  F_{\beta}(U) = \frac{1}{d} \operatorname{Tr}\left[ \left( \rho_{\beta} \otimes \mathbb{I} \right)\left( \sum\limits_{j}^{} \sqrt{\lambda_{j}} U_{j} \otimes W_{j} \right)^{\otimes 2} \mathbb{S}_{AA'} \left( \sum\limits_{k}^{} \sqrt{\lambda_{k}} U^{\dagger}_{k} \otimes W^{\dagger}_{k} \right)^{\otimes 2} \mathbb{S}_{AA'} \right]\\
&= \frac{1}{d} \sum\limits_{jklm}^{} \sqrt{\lambda_{j} \lambda_{k} \lambda_{l} \lambda_{m}} \operatorname{Tr}\left[ \left( \rho_{\beta} \otimes \mathbb{I} \right) \left( U_{j} \otimes W_{j} \otimes U_{k} \otimes W_{k} \right)  \mathbb{S}_{AA'} \left( U^{\dagger}_{l} \otimes W^{\dagger}_{l} \otimes U^{\dagger}_{m} \otimes W^{\dagger}_{m} \right) \mathbb{S}_{AA'} \right]\\
&= \frac{1}{d} \sum\limits_{jklm}^{} \sqrt{\lambda_{j} \lambda_{k} \lambda_{l} \lambda_{m}} \operatorname{Tr}\left[ \left( \rho_{\beta} \otimes \mathbb{I} \right) \left( U_{j} \otimes W_{j} \otimes U_{k} \otimes W_{k} \right) \left( U^{\dagger}_{m} \otimes W^{\dagger}_{l} \otimes U^{\dagger}_{l} \otimes W^{\dagger}_{m} \right) \right],
\end{align*}
where we have used the adjoint action \(\mathbb{S}_{AA'} \left( X_{A}
  \otimes Y_{A'} \right) \mathbb{S}_{AA'} = Y_{A} \otimes
X_{A'}\). Then,
\begin{align}
  F_{\beta}(U) = \sum\limits_{jk}^{} \sqrt{\lambda_{j}} \left( \lambda_{k} \right)^{3/2} \operatorname{Tr}\left[ \rho_{\beta} \left( U_{j} \otimes W_{j} \right) \left( U^{\dagger}_{k} \otimes W^{\dagger}_{k} \right) \right],
\end{align}
where we have used the fact that  \(\left\langle U_{j},U_{k}
\right\rangle = d_{A} \delta_{jk}\) and \(\left\langle W_{j}, W_{k}
\right\rangle = d_{B} \delta_{jk}\) and summed over two of the indices
of the summation above. Now, if \(U\) is maximally entangled (namely,
its eigenstates are maximally entangled) across \(d_{A} = d_{B} =
\sqrt{d}\) and one has \(\sqrt{\lambda_{j}} =
\sqrt{\frac{1}{d_{A}^{2}}}\), then
\begin{align}
F_{\beta}(U) &= \frac{1}{d_{A}^{2}} \operatorname{Tr}\left[ \rho_{\beta} \left( \sum\limits_{j}^{} \sqrt{\lambda_{j}} U_{j} \otimes W_{j} \right) \left( \sum\limits_{k}^{} \sqrt{\lambda_{j}} U_{k} \otimes W_{k} \right) \right] = \frac{1}{d_{A}^{2}} \operatorname{Tr}\left[ \rho_{\beta} U U^{\dagger} \right] = \frac{1}{d_{A}^{2}}.
\end{align}
Therefore, \(G_{\beta}(U) = 1 - F_{\beta}(U) = 1 - \frac{1}{d_{A}^{2}}\) and is independent of \(\beta\).
\subsection*{The dynamical map \(\mathcal{U}_{\beta,t}\) is a quantum operation}
Recall that the Choi-Jamiolkowski (CJ) isomorphism is an isomorphism between linear maps \(\mathcal{E}:
\mathcal{L}(\mathcal{H}) \rightarrow \mathcal{L}(\mathcal{K})\) to
matrices \(\rho_{\mathcal{E}} \in \mathcal{L}(\mathcal{H}) \otimes
\mathcal{L}(\mathcal{K})\). Let \(| \phi^{+} \rangle:=
\frac{1}{\sqrt{d}} | j \rangle | j \rangle\) be the
\textit{normalized} maximally entangled state in
\(\mathcal{H}^{\otimes 2}\), then,
\begin{align}
  \rho_{\mathcal{E}}:= \mathcal{E} \otimes \mathcal{I} \left( | \phi^{+} \rangle \langle  \phi^{+} |  \right).
\end{align}
A linear map \(\mathcal{E}\) is CP \(\iff \rho_{\mathcal{E}} \geq 0\). Computing the CJ matrix corresponding to the
linear map \(\mathcal{V}_{\beta}(X) := \exp \left[ -\beta H/4 \right]
X \exp \left[ -\beta H/4 \right] \), we have,
\begin{align}
  \rho_{\mathcal{V}_{\beta}} &= \frac{1}{d} \sum\limits_{j,k=1}^{d} \mathcal{V}_{\beta} \otimes \mathcal{I} \left( | j \rangle \langle  k |  \otimes | j \rangle \langle  k |  \right)\\
&=\frac{1}{d} \sum\limits_{j,k=1}^{d} \exp \left[ -\beta \left( E_{j} + E_{k} \right)/4 \right]   \left( | j \rangle \langle  k |  \otimes | j \rangle \langle  k | \right)   = \frac{\mathcal{Z}(\beta/2)}{d} | \psi(\beta/2) \rangle \langle  \psi(\beta/2) | ,
\end{align}
where in the second equality we have used the expansion of the
TDS \cref{eq:thermofield-double-energy}. Now,
since \(| \psi(\beta/2) \rangle \langle  \psi(\beta/2) | \) is a pure
state (or a rank-\(1\) projector), it is positive semidefinite. Moreover,
\(\frac{\mathcal{Z}(\beta/2)}{d} \geq 0\), therefore,
\(\rho_{\mathcal{V}_{\beta}} \geq 0\). As a result,
\(\mathcal{V}_{\beta}\) is a CP map.

To show that the map \(\mathcal{V}_{\beta}\) is trace-nonincreasing, we have to show that \(\operatorname{Tr}\left[
  \mathcal{V}_{\beta}(\rho) \right] \leq \operatorname{Tr}\left[ \rho
\right] ~~\forall  \rho \in \mathcal{B}(\mathcal{H})\). Namely,
\begin{align}
& \iff \operatorname{Tr}\left[ \mathcal{V}_\beta(\rho) - \rho \right] \leq 0 ~~\forall  \rho,\\
& \iff \operatorname{Tr}\left[ \exp \left[ -\beta H/4 \right] \rho \exp \left[ -\beta H/4 \right] -\rho   \right] \leq 0 ~~\forall \rho\\
& \iff \operatorname{Tr}\left[ \left( \exp \left[ -\beta H/2 \right] - \mathbb{I} \right) \rho  \right] \leq 0 ~~\forall \rho,\\
& \iff \exp \left[ -\beta H/2 \right] \leq \mathbb{I},
\end{align}
where in the last line, we have used the definition of positive
semidefiniteness. Now, assuming \(H \geq 0\), we have, \(\exp \left[
  -\beta H/2 \right] \leq \mathbb{I}\) since \(\exp \left[ -\beta
  E_{j}/2 \right] \leq 1\) if \(\beta, E_{j} \geq 0 ~~\forall
j\). Notice that, the matrix \(\rho_{\mathcal{V}_{\beta}}\) is
\textit{subnormalized} with \(\operatorname{Tr}\left[
  \rho_{\mathcal{V}_{\beta}} \right] = \frac{\mathcal{Z}(\beta/2)}{d}
\leq 1\). Therefore, the dynamical map, \(\mathcal{V}_{\beta}\) is a CP and
trace-nonincreasing linear map, that is, it is a physical quantum
operation \cite{nielsen_quantum_2010} and we can think of \(\rho_{\mathcal{V}_{\beta}}\) as a subnormalized density
matrix corresponding to a quantum process
\cite{PhysRevA.97.012127,PhysRevLett.120.040405}.

\subsection*{Equilibration value of NRC-PS}
To compute the equilibration value, we need to evaluate the
\(R\)-matrix. Recall that \(H_{\mathrm{NRC\text{-}PS}}:=
  \sum\limits_{j,k=1}^{d_{A},d_{B}} E_{j,k} | \phi_{j}^{(A)} \rangle
  \langle  \phi_{j}^{(A)} | \otimes | \phi_{k}^{(B)} \rangle
  \langle  \phi_{k}^{(B)} | \), with the additional assumption that the spectrum \(\{ E_{j,k}
  \}_{j,k}\) satisfies NRC. Define the index \(\alpha \equiv (j,k)\)
  with \(\alpha \in \{ 1,2, \cdots,d \}\) and \(j \in \{ 1,2, \cdot,
  d_{A} \}, k \in \{ 1,2, \cdots, d_{B} \}\). Then, \(\rho_{\alpha}^A = \operatorname{Tr}_{B}\left[ \rho_{\alpha}
  \right] = \operatorname{Tr}_{B}\left[ \rho_{jk} \right] = |
  \phi^{A}_{j} \rangle \langle  \phi^{A}_{j} | \), that is for all
  paired indices \((j,k)\) and \((j,k')\) the reduced states are the
  same. For the \(R\)-matrix, consider \(\alpha = (j,k)\) and \(\beta =
  (l,m)\), we have, \(R^{A}_{\alpha,\beta} = \left\langle
    \rho_{\alpha}^A,\rho^{A}_\beta \right\rangle =
  \delta_{j,l}\). Similarly, we have,  \(\rho_{\alpha}^B = \operatorname{Tr}_{A}\left[ \rho_{\alpha}
  \right] = \operatorname{Tr}_{A}\left[ \rho_{jk} \right] = |
  \phi^{B}_{k} \rangle \langle  \phi^{B}_{k} | \) and
  \(R^{B}_{\alpha,\beta} = \delta_{k,m}\).

We are now ready to evaluate
\(\overline{G^{(r)}_{\beta}(t)}\). Notice that
\begin{align}
  &\sum\limits_{\alpha,\beta=1}^{d} \exp \left[ -\beta/2 \left(
    E_{\alpha} + E_{\beta} \right) \right] \left| R_{\alpha,\beta}^A
  \right|^{2} = \sum\limits_{j,k,l,m=1}^{d_{A},d_{B}} \exp \left[ -\beta/2 \left( E_{j,k} + E_{l,m} \right) \right] \left|  \delta_{j,l} \right|^{2}\\
  &=  \sum\limits_{j,k,m=1}^{d_{A},d_{B},d_{B}} \exp \left[ -\beta/2 \left( E_{j,k} + E_{j,m} \right) \right] = \sum\limits_{j=1}^{d_{A}} \left( \sum\limits_{k=1}^{d_{B}} \exp \left[ -\beta E_{j,k}/2\right]  \right)^{2}.
\end{align}
Defining, \(\Theta_{j}^A :=  \left( \sum\limits_{k=1}^{d_{B}} \exp
  \left[ -\beta E_{j,k}/2\right]  \right)^{2}\), we have, \(\sum\limits_{\alpha,\beta=1}^{d} \exp \left[ -\beta/2 \left(
    E_{\alpha} + E_{\beta} \right) \right] \left| R_{\alpha,\beta}^A
\right|^{2} = \sum\limits_{j=1}^{d_{A}} \Theta_{j}^{A}\). Similar
algebraic manipulations prove the final result,
\begin{align}
  \overline{G^{(r)}_{\beta}(t)} = \frac{1}{d\mathcal{Z}(\beta)} \left( \sum\limits_{j=1}^{d_{A}} \Theta_{j}^{A} + \sum\limits_{k=1}^{d_{B}} \Theta_{k}^{B} - \mathcal{Z}(\beta)\right).
\end{align}

This can then be rewritten in the form in the main text, namely,
\begin{align}
\overline{G^{(r)}_\beta(t) }|_{\mathrm{NRC-PS}}=\frac{1}{d}\left( \frac{ \|{ \mathbf{p}}^A(\beta/2)\|^2+\| {\mathbf{p}}^B(\beta/2)\|^2}{\| {\mathbf{p}}(\beta/2)\|^2}-1 \right),
\end{align}
where the probability vector ${ \mathbf{p}}(\beta)$ is defined by the components \(p_j=\frac{e^{-\beta E_j}}{Z(\beta)}\) and ${ \mathbf{p}}^{A/B}(\beta)$ are its marginals.

Moreover, for the disconnected correlator, a similar calculation shows
that,
\begin{align}
G^{(d)}_{\beta} = \frac{1}{d\mathcal{Z}(\beta)^{2}} \left( \sum\limits_{j=1}^{d_{A}} \Theta_{j}^{A} \right) \left(  \sum\limits_{k=1}^{d_{B}} \Theta_{k}^{B}\right).
\end{align}

\subsection*{Numerical details}
The \(R^{2}\) value for all linear fits was \(\gtrapprox 0.99\) for
all data and hence we do not report the numbers here.
  \begin{table}[h]
  \centering
  \begin{tabular}{||c |c| c| c||}
 \hline
 Model & \(\beta = 0\) & \(\beta = 1\) & \(\beta = \infty\) \\ [0.5ex] 
 \hline\hline
TFIM integrable & 0.523785 & 0.0341617 &  -0.525957\\ 
 \hline
NRC-PS & 0.971448 & 0.83213 & 1.00 \\
 \hline
Anderson & 1.13615 & 0.742852 & \(6.79567 \times 10^{-7}\) \\
 \hline
MBL & 0.700658 & 0.213039 & -0.012799 \\
 \hline
TFIM chaotic & 1.60648 & 0.327334 & -0.00502613\\
 \hline
GUE & 1.12868 & 2.56073 & 2.10215 \\ [1ex] 
    \hline
\end{tabular}
\caption{The \(\log_{2}(\alpha)\) for various
  Hamiltonian models at \(\beta = 0,1,\infty\), given the Ansatz \(\overline{G^{(r)}_{\beta}(t)} = \alpha
d^{-\gamma}\).}
\label{table-parametervalues-appendix}
\end{table}

\end{document}